\documentclass[structabstract]{aa}  
\usepackage{float}                                   
\usepackage{amsmath}
\usepackage{natbib}
\usepackage{lscape}
\usepackage{fixltx2e}
\usepackage{rotating}
\usepackage[none,bottom,light,timestamp]{draftcopy}
\usepackage{graphicx}
\usepackage[draft]{hyperref}
\usepackage{txfonts}
\usepackage{xcolor}
\usepackage{hyperref}
\usepackage{mathrsfs}
\usepackage{bbding}
\usepackage{verbatim}
\usepackage{color}
\usepackage{tabularx,ragged2e,booktabs,caption}
\newcolumntype{C}[1]{>{\Centering}m{#1}}

 \def\mso{\,\mathrm{M}_\odot}
 \def\rso{\,{\rm R}_\odot}
 \def\lso{\,{\rm L}_\odot}
 \newcommand{\gcm}{\,{\rm g}\,{\rm cm}^{-3}}
 \def\hrd{HR diagram\,}
 \def\kms{\, {\rm km}\, {\rm s}^{-1}}
 
 \def\llso{\log\, (L/{\rm L}_\odot) \,}
 \def\msoy{\, \mso~{\rm yr}^{-1}}

 \def\teff{T_{\rm eff}}
 \def\Ge{\Gamma_{\rm e}}
\def\gammax{\ensuremath{\Gamma_{\rm max}}}
\def\shrL{\ensuremath{\mathscr{L} / \mathscr{L}_{\odot}}}
\def\inflation{\ensuremath{\Delta r/r_{\rm core}\,}}
\def\kmax{\ensuremath{\kappa_{\rm Fe}^{\rm max}\,}}
\def\izw{I\,Zw18\,}
\def\pop3{Pop\,III}
\def\zmw{\ensuremath{Z_{\rm MW}\,}}
\def\zlmc{\ensuremath{Z_{\rm LMC}\,}}
\def\zsmc{\ensuremath{Z_{\rm SMC}\,}}
\def\zizw{\ensuremath{Z_{\rm \izw}}}
\def\zpop3{\ensuremath{Z_{\rm PopIII}\,}}
\def\menv{\ensuremath{M_{\rm env}\,}}

\newcommand{\deb}[1]{\mathrm{#1}}

\begin{document}

   \title{The metallicity dependence of envelope inflation in massive stars}


   \author{D. Sanyal \inst{1}
          \thanks{e-mail: dsanyal@astro.uni-bonn.de}  
         \and
          N. Langer \inst{1}
         \and
          D. Sz\'ecsi \inst{1,3}
          \and
          S.-C Yoon \inst{2}
             \and
           L. Grassitelli \inst{1}
          }

   \institute{Argelander-Insitut f\"ur Astronomie, Universit\"at Bonn, Auf
              dem H\"ugel 71, 53121 Bonn, Germany
         \and
         Department of Physics and Astronomy, Seoul National University, Seoul, 151-747, Korea
         \and
         Astronomical Institute of the Czech Academy of Sciences, Ond\v{r}ejov,
Czech Republic
             }

   \date{Received September 2016; accepted November 18, 2016}

 
  \abstract
   {Recently it has been found that models of massive stars reach the Eddington limit in their
interior, which leads to dilute extended envelopes.
   }
   {We perform a comparative study of the envelope properties of massive stars at different metallicities, with the aim to establish the impact of the stellar metallicity on the effect of envelope inflation.}
   {We analyse published grids of core-hydrogen burning massive star models computed with metallicities appropriate for massive stars in the Milky Way, the LMC and the SMC, the very metal poor dwarf galaxy I\,Zwicky\,18, and for metal-free chemical composition.}
   {Stellar models of all the investigated metallicities reach and exceed the Eddington limit in their interior, aided by the opacity peaks of iron, helium and hydrogen, and consequently develop inflated envelopes. Envelope inflation leads to a redward bending of the zero-age main sequence and a broadening of the main sequence band in the upper part of the Hertzsprung-Russell diagram. We derive the limiting $L/M$-values as function of the stellar surface temperature above which inflation occurs, and find them to be larger for lower metallicity. While Galactic models show inflation above $\sim29\mso$, the corresponding mass limit for Population\,III stars is $\sim150\mso$. While the masses of the inflated envelopes are generally small, we find that they can reach $1-100\mso$ in models with effective temperatures below $\sim8000$\,K, with higher masses reached by models of lower metallicity.
}  
   {Envelope inflation is expected to occur in sufficiently massive stars at all metallicities, and is expected to lead to rapidly growing pulsations, high macroturbulent velocities, and might well be related to the unexplained variability observed in Luminous Blue Variables like S\,Doradus and $\eta$\,Carina.
     }


    \authorrunning{D. Sanyal et al.}
   \maketitle
%

\section{Introduction}\label{sec:intro}

Massive stars, although rare, are cosmic engines in the universe. They drive the dynamical and chemical evolution of galaxies with their strong stellar winds, high luminosities and spectacular explosions. The earliest massive stars, i.e. the metal-free Population III stars may have played a major role in the reionisation of the universe \citep{bromm2009}.  Furthermore, massive stars in low metallicity environments are primary candidates for progenitors of long duration gamma ray bursts \citep{yl05,wh2006}, pair instability supernovae \citep{hw2002,langer2007} and superluminous supernovae \citep{quimby2013,inserra2014}. Accurate models of massive stars across a wide range of metallicities are therefore needed to facilitate comparisons with the available observational data \citep{mm2012,langer2012}.

Recently evidence has accumulated that stars more massive than the often quoted upper mass limit of $\sim150\mso$ \citep{figer2005} exist in the local universe. For example, \citet{crowther2010} estimated present day masses of up to $260\mso$ for several stars in the Tarantula nebula of the LMC. Furthermore, in the recently concluded VLT-FLAMES Tarantula Survey of massive stars in the LMC \citep{evans2011}, \citet{bestenlehner2014} identified three stars with initial mass estimates above $150\mso$. Therefore  models of very massive stars with up-to-date physics have become increasingly relevant.

The mass-luminosity ($M-L$) relation for main-sequence stars, $L\propto M^{\alpha}$, has $\alpha>1$.  However, for constant opacity, $\alpha\rightarrow 1$ for $M\rightarrow \infty$ \citep{kw90}. Therefore one might wonder whether the Eddington limit, which is proportional to $L/M$, is ever reached by stars. 

The classical Eddington limit, that is proportional to the electron-scattering opacity and the star's $L/M$ ratio, is not reached for stars below $\sim10^5\mso$  for Solar composition \citep{kato85,kato86}. It was shown by \citet{sanyal2015} that even when the Rosseland mean opacity is considered, stars below $\sim500\mso$ do not reach the Eddington limit at their surface.  But, when the Eddington limit is defined in the stellar interior \citep{langer97}, \citet{sanyal2015} showed that  main-sequence models with LMC composition reach and exceed the Eddington limit at masses $M\gtrsim40\mso$.  Such stellar models, instead of developing a strong outflow, re-adjust their structure such that a dilute and extended envelope is produced, a process which is called envelope inflation. This effect was earlier pointed out by \citet{ishii99} for zero-age main-sequence models, and by \citet{petrovic_2006} and \citet{graefener_2012}  for helium star models. As a result of such an envelope structure the surface temperatures of these models are much lower than they would have been without this effect, which has consequences for the further evolution of the stars \citep{koehler2015}. Indeed, the distribution of OB stars in our Galaxy shows many stars with masses $>30\mso$ in the effective temperature range  $10\,000-30\,000\,$K, and it has been suggested that these stars are affected by envelope inflation \citep{castro_2014}.

The metallicity ($Z$) of a star affects many of its physical properties  like the wind mass-loss rate, opacity and the equation of state.  The OPAL opacity tables from \citet{ir92} introduced an opacity peak at a temperature of $T\sim2\times 10^5\,$K caused by bound-bound and bound-free transitions of iron-group elements.  This so-called Fe-bump opacity drastically changed the envelope structure of stellar models \citep{stot_chin1993} and even new pulsational instability strips were discovered in the Hertzsprung-Russell diagram \citep{pamyatnykh1999}. Since the Eddington factor depends on opacity, the Fe-bump which is a function of $Z$  plays a major role in determining the extent of envelope inflation in a massive star model.

In this paper, we extend the study by \citet{sanyal2015} over a wide range of metallicities from Galactic to metal-free, and investigate the properties of the stellar models in the context of the Eddington limit and envelope inflation. Section \ref{subsec:models} presents an overview of the grids of models used in this study while in Sec.\,\ref{sec:edd_limit} and \ref{sec:inflation} respectively we explain our concept of the Eddington limit and envelope inflation. We discuss how the Eddington limit and envelope inflation change with metallicity in Sec.\,\ref{sec:results}. We  give our conclusions in Sec.\,\ref{sec:conclusions}.

\section{Method}\label{sec:modelling}
\subsection{Stellar models}\label{subsec:models}

The stellar models used in the present study were computed with a one-dimensional hydrodynamic Lagrangian code (BEC) that includes up-to-date input physics including rotation \citep[for details, see ][and references therein]{heger2000,yln06,ines2011,koehler2015}. Grids of models computed with five metallicities were used, appropriate for the Milky Way (MW), LMC, SMC, I\,Zwicky\,18 (\izw) and for Population III (Pop\,III) stars. The MW, the LMC and the SMC models are published in \citet{ines2011} and \citet{koehler2015}, whereas the \izw and the Pop\,III  models are from \citet{szecsi2015} and \citet{yoon2012},  respectively. The initial chemical compositions and the initial mass ranges in each of these grids are summarised in Table \ref{tab:init_massrange}. In this paper, we consider only core hydrogen burning models that are either non-rotating or slowly rotating, i.e. with $v_{\rm rot}\le 100\,{\rm km\,s^{-1}}$, $v_{\rm rot}$ being the equatorial rotational velocity at the photosphere.

\begin{table*}[!htb]
  \caption{The initial chemical compositions (in mass fraction), the metallicities and the range of initial masses of the stellar evolutionary sequences that were used in this study.} \label{tab:init_massrange}
\begin{center}
\begin{tabular}{C{2.5 cm} | C{2.5 cm} | C{2.5 cm} | C{2.5 cm} | C{2.5 cm} | C{2.5 cm}}
\hline \hline
\noalign{\smallskip} 
 & MW & LMC & SMC & \izw & PopIII \\
\hline 
\noalign{\smallskip} 
  $X_{\rm Fe}$  & $1.02\times10^{-3}$      &  $4.64\times10^{-4}$ &  $2.52\times10^{-4}$  &  $2.52\times10^{-5}$   &  0 \\
  $X_{\rm O}$   & $4.14\times10^{-3}$      &  $2.65\times10^{-3}$ &  $1.14\times10^{-3}$  &   $1.14\times10^{-4}$  &  0 \\
  $X_{\rm He}$  & $0.264$  & $0.256$  & $0.251$   & $0.248$    & $0.240$  \\
  $Z$           & $0.0088$ & $0.0047$ & $0.0021$  & $0.00021$  & $0$  \\
  \hline
  $M_{\rm init}\,(\mso)$      & $3-100$ &  $1-500$&  $5-60 $   & $9-294$    & $10-1000$  \\
  \hline \hline
\end{tabular}
\end{center}
\end{table*}

The standard non-adiabatic mixing length theory \citep[MLT,][]{MLT1958,kw90}  was used to model the energy transport in the convective zones in the stellar interior  with a mixing length parameter of $\alpha=l/H_{\rm p}=1.5$ \citep{lan91},  where $l$ is the mixing length and $H_{p}$ is the pressure scale height.  A discussion of the properties of convection in the inflated envelopes of our LMC models can be found in \citet{sanyal2015}. The parameters for core-convective overshooting ($\alpha=0.335$) and rotationally induced chemical mixing ($f_c=0.0228$, $f_{\mu}=0.1$) were adopted from \citet{ines2011}. Transport of angular momentum  by Spruit-Tayler dynamo \citep{spruit02} was treated  following \citet{petrovic2005}. Radiative opacities from the OPAL tables \citep{ir96}  were used for temperatures above $8000\,$K. For temperatures below $8000\,$K the opacity tables from \citet{AF1994} were used.

\subsection{Mass-loss}
\subsubsection*{MW, LMC, SMC and \izw models:} For stellar models with effective temperatures higher than $22\,000\,$K, 
the mass-loss prescription from \citet{vink2001} was employed to account for the winds of  O- and B-type stars. The mass-loss rate prescription from \citet{nieuwenhuijzen1990} was used  at effective temperatures less than $22\,000$ K, if the  \citet{nieuwenhuijzen1990} mass-loss rate exceeded that of \citet{vink2001}.  In the Wolf-Rayet (WR) evolutionary phases, i.e. when the surface helium  mass fraction ($Y_{\rm s}$) is greater than $70\%$, the empirical mass-loss rates from \citet{hamann95} was used, scaled down by a factor of 10 \citep{yln06}. For $0.4\leq Y_{\rm s}\leq0.7$, a  linear interpolation between the \citet{vink2000,vink2001} mass-loss rate and the  \citet{hamann95} mass-loss rate reduced by a factor of 10, was used.

\subsubsection*{Pop\,III models:} 
For metal-free hot stars, a very low mass-loss rate of $10^{-14}\msoy$ is predicted near the classical Eddington limit \citep{marigo2003,krticka2006,yoon2012}. Therefore, in the \pop3 grid stellar wind mass-loss rates were applied only if there was any surface enrichment of CNO elements by rotational mixing. Hence the non-rotating models practically did not suffer from any mass-loss over their lifetime. For rotating stars, the mass-loss prescriptions from \citet{kudritzki89} and \citet{nieuwenhuijzen1990} were used for $T_{\rm eff}>10^4\,$K and $T_{\rm eff}<10^4\,$K respectively, with a metallicity scaling of $Z^{0.69}$. 

\subsubsection*{Rotationally enhanced mass-loss}
The effect of rotationally enhanced mass-loss \citep{friend1986,langer97} is treated in our models as, 
\begin{equation}\label{eq:mdot_enhanced}
 \dot{M}(v_{\rm rot}) = \dot{M}(v_{\rm rot}=0)\left(\frac{1}{1-\Omega}\right)^{0.43},
\end{equation}
where 
\begin{equation}
 \Omega = \frac{v_{\rm rot}}{v_{\rm crit}}\,\,\, {\rm and}\,\,\, v_{\rm crit} = \sqrt{\frac{GM}{R}(1-\Gamma_{\rm avg})}.
\end{equation}
Here$v$ $\Gamma_{\rm avg}$ is the Eddington factor averaged over the region with optical depth between $2/3$ and $100$.  The enhancement of the mass-loss rate is limited by the thermal timescale mass-loss rate of the star to avoid the singularity in Eq.\,\eqref{eq:mdot_enhanced} as $v$ approaches $v_{\rm crit}$ \citep{yoon2012}. Note that for the models analysed in this paper the enhancement to the mass-loss rate is negligible.

\subsection{Additional models}
In the evolutionary sequences computed by \citet{ines2011} and \citet{koehler2015}, the data regarding the structure of a stellar model is stored for every $50^{\rm th}$ computed model, i.e., at non-regular time intervals since the time steps are not uniform along the evolution.  For the \izw sequences, every $250^{\rm th}$ model is stored. In order to have a higher model density in certain parts of the \hrd for the present study, several evolutionary tracks (without rotation) were re-computed with the same input parameters, as summarised in Table \ref{tab:masses}.
\begin{table}[!htb]
  \caption{The initial masses (in units of $\mso$) of the non-rotating evolutionary sequences that were re-computed for this study. } \label{tab:masses}
\begin{center}
\begin{tabular}{C{1.5cm} | C{1.5cm} | C{1.5cm} | C{1.5cm}}
\hline \hline
\noalign{\smallskip} 
MW & LMC & SMC & \izw \\
\hline 
\noalign{\smallskip} 
  40 & 40 & 50 & 100\\
  50 & 60 & 60 & 150\\
  60 & 70 & 80 & 196\\
  80 & 100 &    &\\
  \hline \hline
\end{tabular}
\end{center}
\end{table}
\section{The Eddington Limit}\label{sec:edd_limit}

Conventionally, a star is considered to be at the Eddington limit when its luminosity $L$ equals its Eddington luminosity ($L_{\rm Edd}$), defined as the condition when the radiative acceleration ($g_{\rm rad}$) balances the gravitational acceleration ($g$) at the stellar surface. The radiative acceleration  is proportional to  the stellar luminosity and the opacity $\kappa$. Considering electron scattering as the only source of opacity, i.e., $\kappa=\kappa_e$, the classical Eddington factor is defined as
\begin{equation}\label{eq:gamma_e}
 \Ge = \frac{g_{\rm rad}}{g}=\frac{\kappa_e L}{4\pi c G M},
\end{equation}
where the physical constants have their usual meaning. Using this definition, stellar models reach the Eddington limit only at masses $M\gtrsim 10^5\,\mso$ \citep{kato86}. 

When the Rosseland mean opacities are used, the LMC models from \citet{koehler2015} do not reach the Eddington limit at their surface, even at $500\,\mso$ \citep{sanyal2015}. However, one can also define an Eddington factor locally \citep{langer97} such that 
\begin{equation}\label{eq:gamma}
 \Gamma(r):=\frac{L_{\mathrm{rad}}(r)}{L_{\mathrm{Edd}}(r)}=\frac{\kappa(r) L_{\rm rad}(r)}{4\pi c G M(r)},
\end{equation}
where $M(r)$ is the Lagrangian mass coordinate, $\kappa(r)$ is the Rosseland mean opacity and $L(r)$ is the local luminosity. Since the convective luminosity does not contribute to the radiative acceleration, it is not considered in Eq.\,\eqref{eq:gamma}. Using this definition, core-hydrogen burning LMC models with masses as low as $\sim 40\mso$ reach the Eddington limit in their interior \citep{sanyal2015}. However, instead of a super-Eddington outflow, a hydrostatic structure with an extended envelope  is obtained (cf.\, Sec.\, \ref{sec:inflation}), often associated with a density inversion \citep[cf.\,Fig.\,9 in ][]{sanyal2015}. It has been argued in the literature \citep{langer97,sanyal2015} that the concept of the Eddington limit as a stability criterion does not apply in the stellar interior. If not explicitly stated otherwise, we will use the definition in Eq.\,\ref{eq:gamma} for the Eddington factor in the rest of the paper.

Since the Rosseland mean opacity $\kappa$ depends on the chemical composition and hence the metallicity, the Eddington factor is also expected to be a function of metallicity. \citet{sanyal2015} showed that in the LMC models from \citet{koehler2015}, the opacity peaks caused by the partial ionisation of iron group elements,  helium and hydrogen shape the profile of the Eddington factor  inside a massive star model. In this paper we  investigate how the  metallicity  influences the Eddington factor in the stellar interior. 

\section{Envelope inflation}\label{sec:inflation}
\begin{figure}
\centering
\mbox{\includegraphics[width=6.5cm, angle=-90]{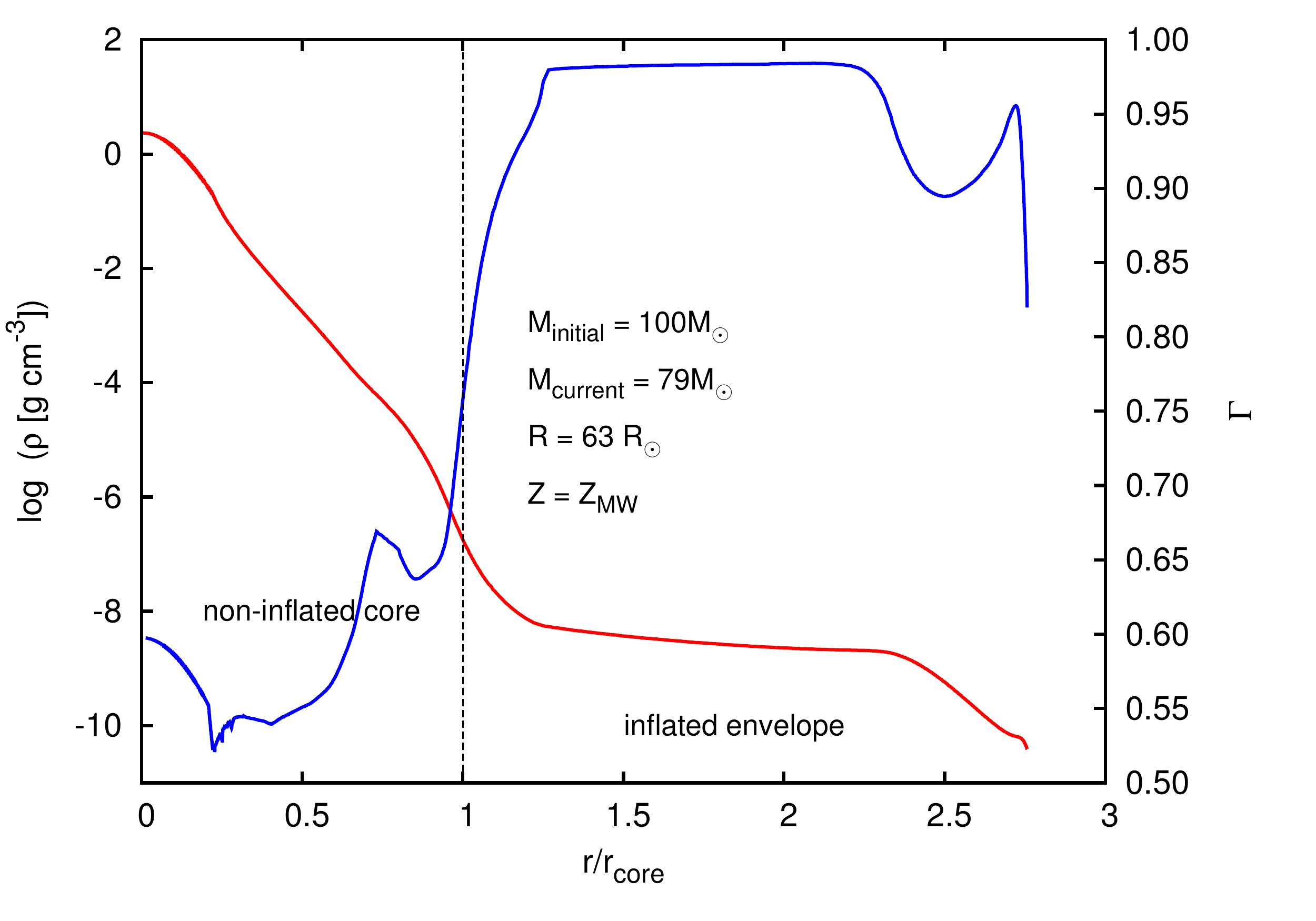}}
   \caption[]{Density profile of a $79\mso$ solar metallicity core-hydrogen burning model showing the inflated envelope. The model had an initial mass of $100\mso$. The radial co-ordinate has been scaled with the core radius $r_{\rm core}$. The blue line (refer to the right Y-axis) shows the run of the Eddington factor $\Gamma$ (Eq. \ref{eq:gamma}) in the interior of the model. The Eddington factor gets close to unity in the inflated region.
     } 
\label{fig:inflation_eg}
\end{figure}

As already mentioned above, stellar models reaching the Eddington limit in their interior have extended and dilute envelopes \citep{ishii99, graefener_2012,koehler2015,sanyal2015}. When the Eddington limit is reached in the interior, either $L_{\rm rad}$ or the opacity $\kappa$ needs to be reduced. But, since energy transport by convection may be inefficient because of low densities,  $L_{\rm rad}$ can not be significantly reduced in this case, and hence the opacity needs to decrease via a further drop in density. As a result an inflated envelope develops such that $\Gamma\approx 1$ is maintained across the inflated region.  An example of the density structure of such an inflated stellar model is shown in Fig.~\ref{fig:inflation_eg} where the region with a steeply declining density profile is referred to as the non-inflated core and the region with  a relatively flat density profile is referred to as the inflated envelope. Whereas the core radius ($r_{\rm core}$) of this model is $25.2\,\rso$, the extent of the inflated envelope is about $1.7$ times that of $r_{\rm core}$. The profile of the Eddington factor shows that in the core it is $\Gamma<1$, but in the envelope $\Gamma\approx 1$. At the surface of the star, $\Gamma$ drops to $0.82$.

To our knowledge, an analytical criterion for inflation is not available in the literature \citep[however, see Sec.\,3.4 and 3.5 in ][]{owocki2015} for models at various evolutionary stages, i.e from  hot WR stars \citep{petrovic_2006} to cool core-hydrogen burning red supergiants \citep{graefener_2012,sanyal2015,szecsi2015}. We note that envelope inflation is different from the formation of extended envelopes in classical red supergiants, as those are post main-sequence stars with a shell-burning source that is responsible for the envelope expansion, through the so-called mirror principle \citep{kw90}. In this study we follow \citet{sanyal2015} to determine whether a stellar model is inflated or not. Since inflation is related to high values of the Eddington factor, which in turn implies dominance of radiation pressure over gas pressure, the inflated region must have a small enough value of $\beta$, where $\beta$ is the ratio of the gas pressure ($P_{\rm gas}$) to the total pressure ($P$).  For all the model grids except the \pop3 grid, we therefore adopt a threshold value of $\beta=0.15$ to identify the base of the inflated envelope $r_{\rm core}$ in a stellar model, in accordance with \citet{sanyal2015}. For the $M_{\rm init}=1000\mso$ sequences in the \pop3 grid we use a lower threshold $\beta=0.1$, because the mass averaged value of $\beta$ ( and the $\beta$-value in the centre) drops below $0.15$ for some models in these sequences, as expected for models with extremely high luminosities. For all other model sequences of lower masses in the \pop3 grid, we use $\beta=0.15$. We  motivate our choice of using  a threshold value of $\beta$  to identify the occurrence of an inflated envelope below.

The equation of hydrostatic equilibrium inside a star is given as
\begin{equation}\label{eq:HE_1}
\frac{dP}{dr} = -\rho g,
\end{equation}
where $\rho$ is the density at $r$. Writing $P=P_{\rm gas} + P_{\rm rad}$ where the radiation pressure is $P_{\rm rad} = \frac{1}{3} aT^4$, it is, 
\begin{equation} 
 - \frac{1}{\rho g}\frac{dP_{\rm rad}}{dr}  = \Gamma,
 \end{equation} 
 and
\begin{equation} \label{eq:HE_2}
- \frac{1}{\rho g}\frac{dP_{\rm gas}}{dr} = 1 - \Gamma.
 \end{equation}
 Therefore, $\Gamma\rightarrow 1$ leads to a vanishing gas pressure gradient and $\Gamma>1$ in the hydrostatic stellar interior merely implies a positive gas pressure gradient \citep{joss73,mesa2013}.
Using Eq.\,\eqref{eq:HE_1} in Eq.\,\eqref{eq:HE_2}, we get
\begin{align}
 \frac{dP_{\rm gas}}{dP} &=  \frac{d(\beta P)}{dP}  \\
 & = \beta + P\frac{d\beta}{dP}\\
 & = 1 - \Gamma. 
\end{align}
Locally, for $\beta$ either constant or slowly varying, which is generally true in inflated envelopes \citep[][cf.\,Fig.\,\ref{fig:inflation_eg}]{sanyal2015}, it is 
\begin{equation}\label{eq:gamma_beta}
\Gamma \simeq 1 - \beta.
\end{equation}
 Therefore $\Gamma\lesssim 1$ implies a low value of $\beta$. Since we adopted $\beta=0.15$ as the inflation criterion, we expect to find inflated models with $\Gamma>0.85$, and this is indeed the case \citep{sanyal2015}. Equation \ref{eq:gamma_beta} was earlier arrived at by \citet{graefener_2012} and the validity of the assumption of a constant $\beta$ in the inflated envelope was shown in studies by \citet{graefener_2012} and \citet{sanyal2015}.

To show that Eq.\,\eqref{eq:gamma_beta} is consistent with the occurrence of a flat density profile,  Eq.\,\eqref{eq:HE_2} can also be written as
 \begin{equation}\label{eq:HE_3}
  \Gamma - 1  = \frac{1}{\rho g} \frac{d}{dr}\left(\frac{R\rho T}{\mu}\right).
 \end{equation}
Rearranging Eq.\,\eqref{eq:HE_3} and dividing by $\rho T$ on both sides (assuming a constant $\mu$), we get
\begin{align*}
 \frac{1}{H_\rho}:=\frac{d\ln \rho}{dr} & = \frac{\mu}{RT}(\Gamma - 1)g - \frac{d\ln T}{dr}\\
 & = \frac{g\mu}{RT}\left( \Gamma -1 +  \nabla\beta \right),
\end{align*}
where  $H_\rho$ is the density scale height and $\nabla$ is the temperature gradient defined as $\nabla:= \frac{d\ln T}{d\ln P}$. Substituting $\beta=1-\Gamma$ in the above expression, we obtain
\begin{equation} \label{eq:H_rho}
 \frac{1}{H_\rho} = \frac{\mu g}{RT}(\nabla + 1)(\Gamma - 1),
\end{equation}
which implies that $H_{\rho}\rightarrow \infty$ as $\Gamma \rightarrow1$. Therefore, when $\Gamma(r)$ is close to unity, the density scale height becomes very large and leads to an extended, flat density profile which we identify as a signature of inflation. We further note that a vanishing density gradient, i.e., $\frac{d\rho}{dr}=0$, is equivalent with the condition $V=\frac{d\ln P}{d\ln r} = (GM_r/r)/(P/\rho)=1$.

Quantitatively, we define inflation in a stellar model as $\Delta r/r_{\deb{core}}:=(R_{\star}-r_{\mathrm{core}})/r_{\mathrm{core}}$, where $R_{\star}$ is the stellar radius and $r_{\rm core}$ is the radial co-ordinate where the $\beta$ value drops below $0.15$ for the first time in the stellar interior. Since there is some arbitrariness in our inflation criterion only those models for which our criterion predicts \inflation $>0.05$ are considered  to be inflated.

\citet{langer97} showed that if the Eddington factor is defined in the stellar interior as
\begin{equation}
\Gamma'(r) = \frac{\kappa(r)L(r)}{4 \pi cGm(r)},
\end{equation}
i.e., taking the total luminosity into account, then the Schwarzschild criterion for convective instability can be written in the following form:
\begin{equation}\label{eq:gamma_withconv}
\Gamma'(r) \geq (1 - \beta)\frac{32 -24\beta}{32 -24\beta - \beta^2}.
\end{equation}
For $\beta\ll 1$, the above inequality reduces to
\begin{equation}\label{eq:schwarz_conv}
\Gamma'(r) \geq (1 - \beta).
\end{equation}
Since $\Gamma = \Gamma' \frac{L_{\rm rad}}{L}$ from Eqns.\,\eqref{eq:gamma} and \eqref{eq:gamma_withconv}, the above inequality can be written as
\begin{equation}
\Gamma \geq (1-\beta)\frac{L_{\rm rad}}{L}
\end{equation}
Since $L_{\rm rad}\leq L$ everywhere inside the star, from Eqns. \eqref{eq:gamma_beta} and \eqref{eq:schwarz_conv} we conclude that the inflated envelope will always be convectively unstable. Furthermore, the densities in the sub-surface convection zones of massive stars are low, and convection is strongly non-adiabatic. As a consequence, particularly in the hot models ($\teff\gtrsim15\,000\,$K), the luminosity carried by convection is much smaller than that carried by radiation, and hence $L_{\rm rad}/L\approx 1$, or $\Gamma \approx \Gamma'$.


\section{Results}\label{sec:results}
\subsection{Hertzsprung-Russell (HR) diagram}

\begin{figure*}
\centering
\mbox{\includegraphics[width=12cm,angle=-90]{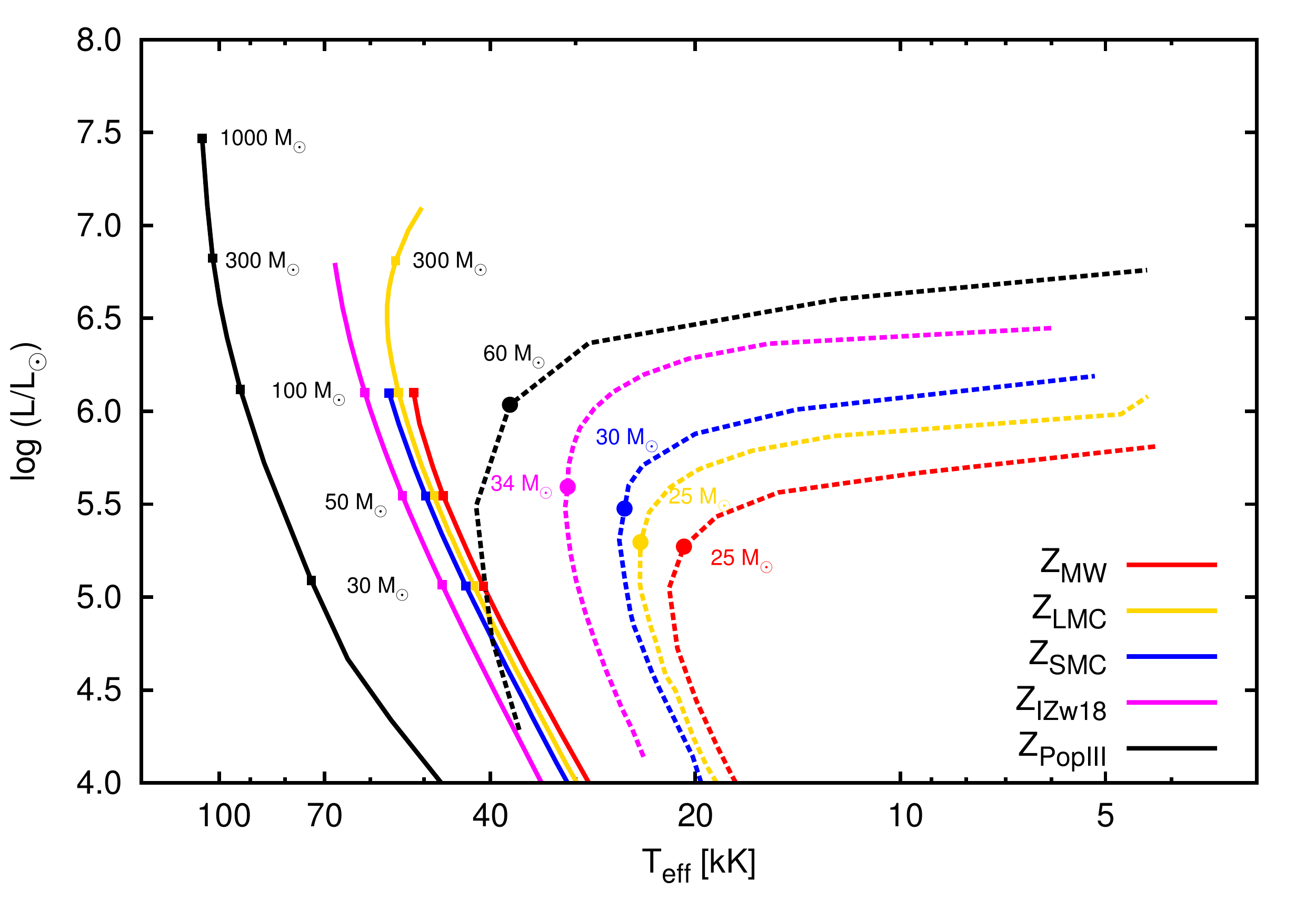}}
   \caption[]{The zero-age main-sequence (solid lines) and the terminal-age main-sequence lines (dotted lines) above $\log\,(L/\lso)>4$ of the analysed stellar models at different metallicities. The masses of some representative models (black squares) are indicated along the ZAMS. The position of the TAMS (in the model grid) where they  bend towards cooler effective temperatures are marked with coloured dots and the the corresponding initial masses of these models are  labelled alongside.} 
\label{fig:ZAMS}
\end{figure*}

The \hrd in Fig.\,\ref{fig:ZAMS} shows the zero-age main-sequence (ZAMS) and the terminal-age main-sequence (TAMS)  of the analysed non-rotating models. The ZAMS is defined by the condition that $3\%$ of hydrogen (by mass) has been burnt whereas the TAMS is defined by the location at  which the models reach the maximum radius over their main-sequence lifetime, i.e., until the  central helium mass fraction is $<0.98$. 
 
The ZAMSs of the different metallicities are located roughly parallel to one another up to $\llso\sim6$ in the \hrd, such that the Pop\,III ZAMS has the highest $\teff$ (and smallest radius) and the MW ZAMS has the lowest $\teff$ (and largest radius) for corresponding model masses.

The LMC ZAMS reaches a maximum $\teff$ of $\sim57\,000\,$K and then starts to bend towards lower values around $\llso\sim6.6$, that corresponds to a mass of $\sim 200\mso$. Above this critical mass the surface temperature of the ZAMS models decrease for increasing mass. The lower the metallicity, the higher is the luminosity and the effective temperature at which the bend is expected to be located \citep{ishii99}. This feature is not seen for all the metallicities in Fig.\,\ref{fig:ZAMS} because the initial mass ranges of the evolutionary sequences do not extend to high enough values (see Table \ref{tab:init_massrange}). Nevertheless one can notice an increase in the slope of the ZAMS's in the upper HR diagram.

Using the latest OPAL opacities, the bending of the ZAMS was earlier investigated by \citet{ishii99}. They computed models with metallicities ranging from $Z=0.1$ to $Z=0.004$ and found that the ZAMS curves redwards at sufficiently high masses for all the metallicities. The Solar metallicity ZAMS in their study had a bend at $M\sim 100\mso$ which is consistent with our results.

The redward curving of the ZAMS is a consequence of envelope inflation of massive luminous stars, as discussed in Sec.\,\ref{sec:inflation}. When the layers in the stellar interior reach the Eddington limit either because of an opacity bump or because of a high $L/M$ ratio,  the high radiation pressure pushes the layers outwards such that density and hence opacity decreases, and the Eddington factor obtains a value $\lesssim1$. 

If convection is efficient then the star does not need to re-adjust its structure, but in the luminous stars discussed here  the low densities in their outer layers imply that convective energy transport within the framework of the standard MLT is not efficient enough to bring down the Eddington factor below one, even though the fraction of the total luminosity carried by convection can exceed $90\%$ in the coolest models \citep[cf.\,][]{sanyal2015}. Therefore the envelope expands giving rise to a core-halo density structure (Fig.\,\ref{fig:inflation_eg}).

\begin{table}[!htb]  
\caption{Model properties (mass, luminosity, effective temperature and the classical Eddington factor) at the points marked by filled dots in Fig.\,\ref{fig:ZAMS}  where the TAMSs bend redwards. } \label{tab:TAMS}
\begin{center}
\begin{tabular}{C{0.9cm} | C{1.3cm} | C{1.3cm} | C{1.2cm} | C{1.0cm} | C{0.7cm}}
\hline \hline
\noalign{\smallskip} 
$Z$ & $M_{\rm init}\,(\mso)$ & $M_{\rm now}\,(\mso)$ & $\llso$ & $\teff$\,[K] & $\Ge$\\
\hline 
\noalign{\smallskip} 
  MW & 25 &  23.6 & 5.27 & 20783 & 0.20\\
  LMC & 25 & 24.4  & 5.29 & 24066 & 0.21\\
  SMC & 30 &  29.3 & 5.48 & 25409 & 0.26\\
  \izw & 34 &  33.8 & 5.59 & 30787 & 0.30\\
  PopIII & 60 & 60.0 & 6.03 & 37417 & 0.47\\

  \hline \hline
\end{tabular}

\end{center}
\end{table}

The redward bend is also present in the TAMS lines of all metallicities. The higher the metallicity, the lower is the luminosity at which the bend occurs, similar to the trend expected for the ZAMS lines. The TAMS however curves redwards at a lower luminosity than the ZAMS. For example, the TAMS for the LMC bends at $\llso\approx 5.3$ whereas the ZAMS bends at $\llso\approx6.6$. From the mass-luminosity  relation for homologous stars, we know that $L\propto \mu^{\beta}$ for a fixed mass, where the exponent $\beta$ lies in the range $\sim1.3\dots 2$ for masses between $100-500\mso$ on the ZAMS \citep[cf. Fig.\,17 in][]{koehler2015} such that higher masses have lower values of $\beta$. Therefore, at the TAMS a model has a higher $L/M$ value than at ZAMS because of a higher mean molecular weight.

The stellar parameters at the points (marked with filled dots in Fig.\,\ref{fig:ZAMS}) where the TAMS  lines bend  towards cooler effective temperatures are noted in Table \ref{tab:TAMS}. This feature indicates the onset of  envelope inflation because below this bend we do not find any TAMS model to be inflated but above the bend we find inflated models. With a decrease in metallicity the opacity in the stellar envelope decreases (cf.\,Sec.\,\ref{subsec:opacity}) and hence $\Gamma\approx1$ can be achieved only with a higher $L/M$ value. Therefore the low-$Z$ TAMS models show envelope inflation at higher luminosities.  The TAMS extends to temperatures below $\sim5000\,$K,  leading to core hydrogen-burning red supergiant models. 
We note that the lowest luminosity at which we identify an inflated model on the TAMS is higher than the luminosity at which the bend is located. This might be related to our ad-hoc criterion for inflation. 

The TAMS lines for the MW, LMC and \izw models bend bluewards above $\llso$ of $5.7$, $6.0$ and $6.8$ respectively \citep{ines2011,koehler2015}. This has not been included in Fig.\,\ref{fig:ZAMS} for the sake of clarity. The blueward bend occurs because the mass-loss rates in this part of the HR diagram are high enough to strip the hydrogen-rich outer layers of the models and produce helium-rich WR models \citep{ines2011,koehler2015}. Once the helium-rich layers are exposed, the mass-loss rates increase even further such that the models evolve towards higher surface temperatures, towards the helium ZAMS.


 \begin{figure*} 
\centering
\mbox{\includegraphics[width=18cm, angle=-90]{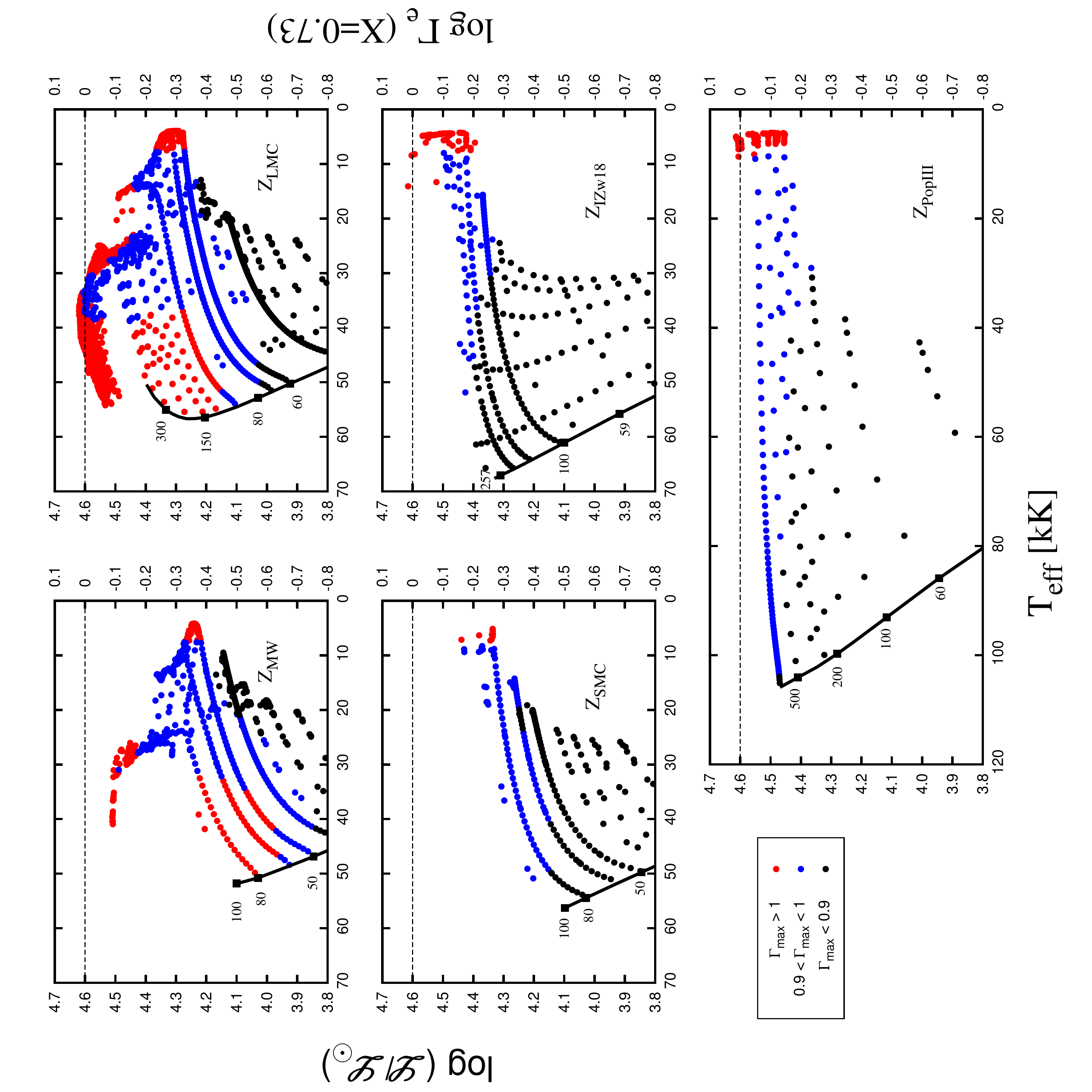}}
\caption[]{Spectroscopic Hertzsprung-Russell diagrams showing the non-rotating and slowly-rotating core-hydrogen burning models in the five grids corresponding to the different metallicities (see labels).   The left Y-axis shows the quantity $\log (\shrL)$ with $\mathscr{L}:=T_{\mathrm{eff}}^4/g$ whereas the right Y-axis shows the corresponding values of $\log (\Ge)$, in all the five panels. The $\Ge$ values are computed at the stellar surface considering electron-scattering opacity with a  hydrogen abundance of $X=0.73$ (completely ionised) and the black dotted line marks the location $\Ge=1$. Note that the assumption of completely ionised hydrogen breaks down for models with $\teff$ below $\approx 10\,000\,$K. Black, blue and red dots correspond to models with $\gammax<0.9$, $0.9<\gammax<1$ and $\gammax>1$, respectively. Only the models with  $\log (\shrL) > 3.8$ have been shown. The black solid line is the ZAMS and the masses of some representative models (in units of $\mso$) have been indicated.}
\label{fig:gammax_Z}
\end{figure*}

\subsection{The spectroscopic HR diagram}
In the spectroscopic HR (sHR) diagram introduced by \citet{langer2014},  
instead of the luminosity the quantity $\mathscr{L}:=T_{\mathrm{eff}}^4/g$ is plotted 
as a function of the effective temperature. The quantity $\mathscr L$ is proportional to $\Ge$ such that 
\begin{equation}
 \Gamma_{\mathrm{e}} = \frac{\kappa_{\deb e} L}{4\pi cGM}=\frac{\kappa_{\deb e} \sigma T_{\mathrm{eff}}^4}{cg}=\frac{\kappa_{\deb e} \sigma}{c}\mathscr{L},
\end{equation}
where  the constants have their usual meaning. Hence for Solar hydrogen abundance,
\begin{equation}
 \log (\Gamma_{\mathrm{e}}) \simeq \log (\shrL)-4.6 .
\end{equation}

Figure \ref{fig:gammax_Z} shows the maximum Eddington factor \gammax\, in the interior of the analysed models for the five  grids. Since the iron bump opacity increases non-linearly (cf.\,Sec.\,\ref{subsec:opacity}) with increasing iron abundance, i.e., with increasing metallicity (Fig.\,\ref{fig:OPAL_Z_BEC}), layers in the stellar interior reach the Eddington limit at a lower $\mathscr{L}$, i.e., at a lower $L/M$. This is demonstrated in the different panels of Fig.\,\ref{fig:gammax_Z}. Whereas we find models with $\gammax>0.9$ for masses as low as $\sim 30\mso$  in the MW grid, the same is achieved at $M \sim 100\mso$ in the Pop\,III grid. Furthermore, an evolutionary model with a  higher initial mass encounters a higher $\gammax$ earlier in its evolution because of its  higher $L/M$ ratio. For example, the $50\mso$ MW sequence starts to develop super-Eddington layers in the midst of its main-sequence life, whereas the $80\mso$ sequence already has $\gammax>1$ on its ZAMS.

In the MW and the LMC grids, there are models with $\gammax>1$ in the $\teff$ range $35-55\,$kK. These models have the Fe opacity peak close to their surface where convective energy transport is inefficient such that $\gammax$ reaches values above one \citep{sanyal2015}. In the temperature range $20-30$ kK but at $\log\,(\shrL)>4.4$, we also find models with $\gammax>1$.  These models are hydrogen-deficient, either because of strong wind mass-loss or because of rotationally-induced mechanical mass-loss in the past.  The super-Eddington layers in these models are caused by the helium opacity bump located close to their surface, coupled with inefficient convection. The models with SMC metallicity or lower do not evolve to have helium-rich envelopes during their main-sequence evolution, at least not in the mass and rotational velocity range considered here. For the  SMC and the \izw metallicity, the Fe-opacity peak, although present, is much weaker compared to the MW and the LMC.  In other words, to reach the same value of $\gammax$ the models with lower metallicity need to have a higher $L/M$ ratio.

\begin{figure}
\centering
\mbox{\includegraphics[width=6.5cm, angle=-90]{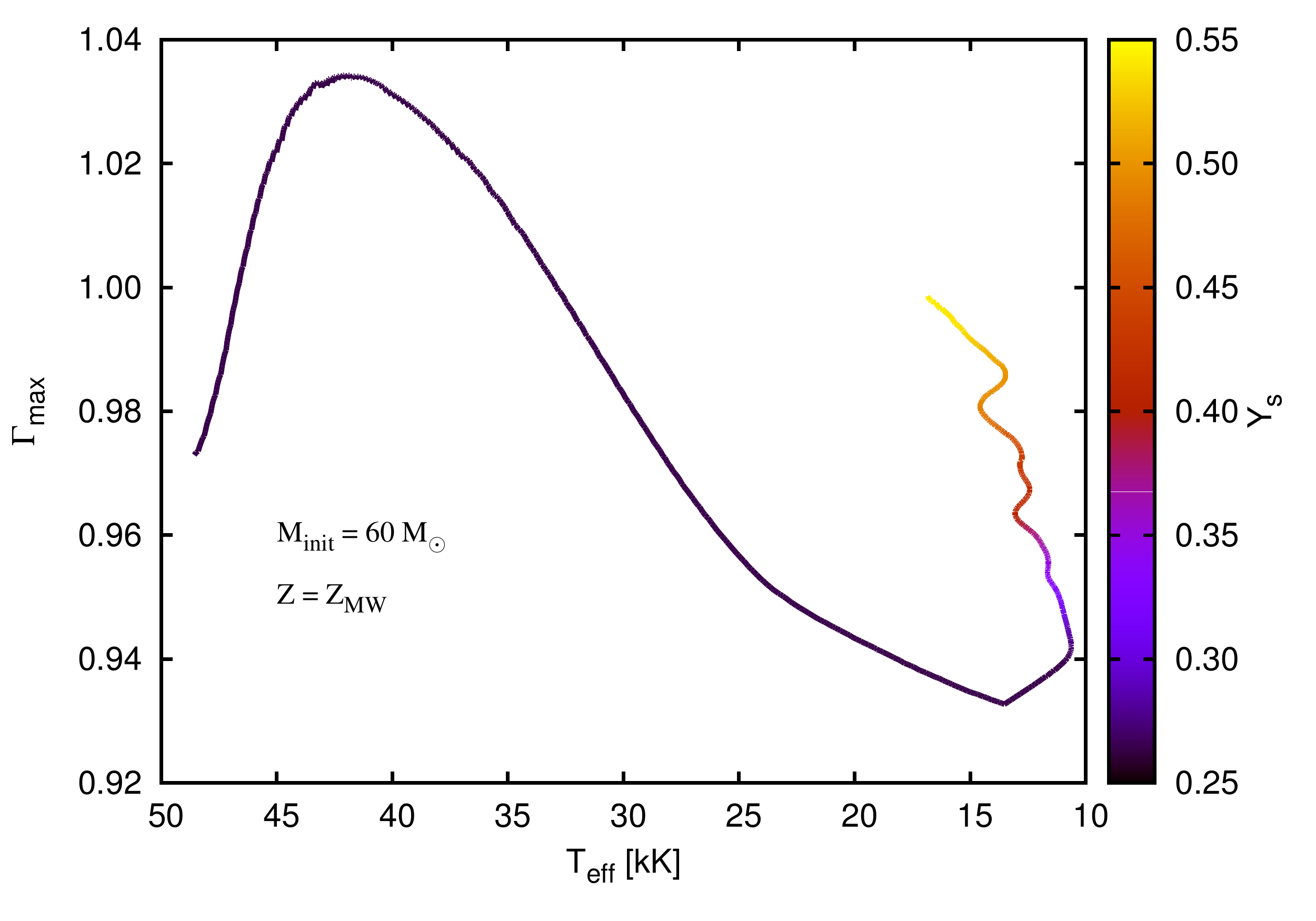}}
\caption[]{Evolution of $\gammax$ with $\teff$ for the $60\mso$ \zmw sequence. The colour depicts the surface helium mass fraction.}
\label{fig:gammax_eg_60msun}
\end{figure}

In the $60\mso$ MW sequence for example, $\gammax$ exceeds unity very close to the ZAMS but at $\teff<32\,$kK $\gammax$ falls below one. This drop in $\gammax$ is explained by relatively efficient convection in the envelope as the Fe-bump moves deeper into the star where densities are relatively higher. The evolution of $\gammax$  versus $\teff$ for the $60\mso$ sequence is shown in Fig.\,\ref{fig:gammax_eg_60msun}.  The increase of $\gammax$ at $\teff<14\,000\,$K is explained by strong mass-loss that increases the $L/M$ ratio and the surface helium abundance. However we note that $\Ge$ increases throughout its main-sequence evolution. A similar trend exists in other evolutionary sequences in Fig.\,\ref{fig:gammax_Z}. 

None of the Pop\,III models in the investigated parameter range have $\gammax>1$ at $\teff>10\,000\,$K. Since the Fe-bump is completely absent, $\Gamma=1$ is never reached.  Neither are these models helium-enriched at their surface because of negligible wind mass-loss that would have increased their $L/M$ ratio. 

The $\log\,\Ge$ values shown on the right Y-axis in Fig.\,\ref{fig:gammax_Z} gives little information about the $\gammax$ in the stellar interior. For example the $80\mso$ MW ZAMS model has super-Eddington layers in its envelope but its $\Ge$ value is only $0.27$. This shows that $\Ge$ is not a good proxy for the true $\Gamma$ while investigating the structure and the stability of massive star envelopes.

At the stellar surface the classical Eddington factor can not exceed unity if hydrostatic equilibrium is to be maintained. Therefore  $\Ge=1$ is an impenetrable upper limit \citep{eddington_1926,langer2014}. However in the LMC grid there are apparently many models with $\Ge >1$ (Fig.\,\ref{fig:gammax_Z}), but their surface  helium mass fraction exceeds $Y_{\rm s}=0.8$ \citep{koehler2015}. The  true $\Ge=1$ for these models therefore is located at a higher $\mathscr{L}$ such that they all lie below it. For example, if $X=0$ the $\Ge=1$ line in Fig.\,\ref{fig:gammax_Z} shifts upwards by $0.24$ dex.

Across all the metallicities, there are models with $\gammax>1$ at $T_{\rm eff}< 9000\,$K. This is because of the opacity peak caused by hydrogen recombination, and hence is not influenced by the metal content in the star.  The $\gammax$ values of these models can be as high as $6$ for the MW models to $\gtrsim 8$ for the Pop\,III models, i.e., in the outer envelope (around the hydrogen recombination temperature) of such a model the luminosity  transported by radiation can be a few times the Eddington luminosity \citep{sanyal2015}. The opacities in the hydrogen recobination zone can be $\sim10$ times that of the Fe-opacity peak.  Hydrostatic equilibrium in these super-Eddington layers is maintained by building up a positive gas pressure gradient and a positive density gradient \citep{joss73,sanyal2015}.  

One might expect that these peculiar structures, coupled with the fact that they are located beyond the observed Humphreys-Davidson (H-D) limit \citep{hd1979} are prone to various instabilities and possibly undergo violent mass-loss episodes such that it prevents them from staying long enough on the cool side of the H-D limit. However, in our hydrodynamic 1-D models we find no sign of a super-Eddington outflow.

\subsection{Dependence of envelope inflation on metallicity}\label{subsec:inflation_Z}

\begin{figure*}[h!]
\centering
\mbox{\includegraphics[width=18cm, angle=-90]{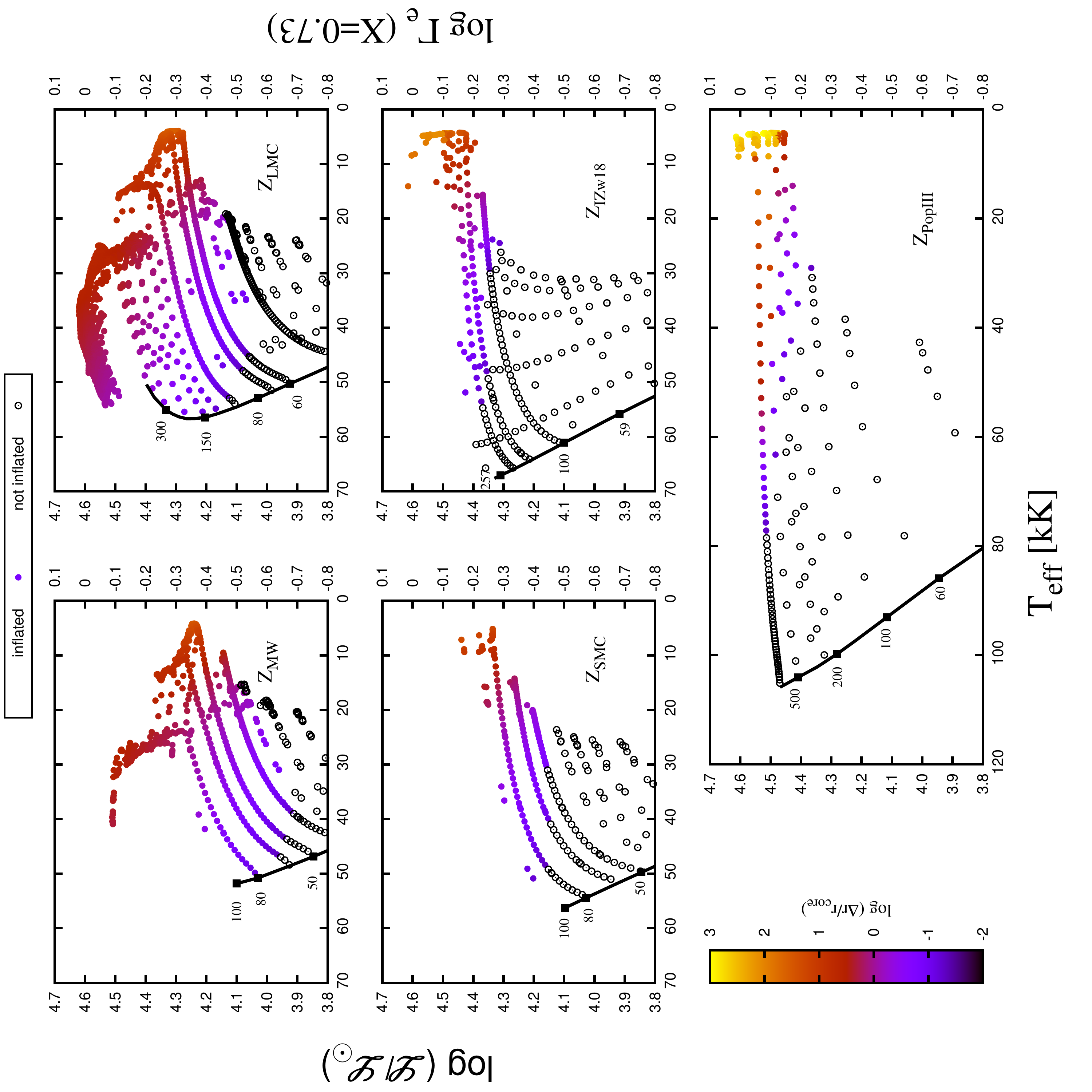}}
   \caption[]{An sHR diagram showing the metallicity dependence of inflation for the analysed models in our grid. The left Y-axis shows the quantity $\log (\shrL)$ whereas the right Y-axis shows the corresponding values of $\log (\Ge)$. The $\Ge$ values are computed assuming a solar hydrogen abundance (completely ionised). Models marked with open black dots are not inflated whereas the coloured dots represent models with inflated envelopes. The colour of the dots indicate the strength of inflation, $\log\,(\inflation)$. The black line is the ZAMS, and the masses of some representative models (in units of $\mso$) are indicated along it.}
\label{fig:inflation_Z}
\end{figure*}

The extent of envelope inflation in the analysed models is summarised in the sHR diagrams in Fig.\,\ref{fig:inflation_Z}. Comparing with Fig.\,\ref{fig:gammax_Z},  we note that barring a few, none of the models with $\gammax<0.9$ are inflated whereas models with $\gammax >1$ all have inflated envelopes. Therefore, as mentioned in Sec.\,\ref{sec:inflation} the occurrence of inflated envelopes is related to models approaching the Eddington limit (as defined by Eq.\,\eqref{eq:gamma}) in their interior. In general, the hotter models are less inflated than the cooler models for a given $\mathscr{L}$, in agreement with the results obtained by \citet{sanyal2015}. This is expected because the effective temperature is strongly affected by inflation. The strongest inflation is found in models with $\teff\lesssim8000\,$K, for all $Z$. The most extreme cases are found in the \izw and the \pop3 models, where \inflation can go up to a few hundred.


The Eddington limit is either approached  with large opacities or with a high $L/M$ ratio. Models with lower metal abundances, i.e. with a weaker Fe-opacity bump, need to attain a higher $L/M$ ratio, or a higher $\Ge$, to reach the Eddington limit and inflate their envelopes.  Therefore inflation starts at higher $\mathscr{L}$ as $Z$ goes down. However, we reiterate that reaching the Eddington limit is a sufficient, but not a necessary condition for envelope inflation to occur (cf.\,Eqns.\,\eqref{eq:gamma_beta} and \eqref{eq:H_rho}) because the gas pressure gradient might also contribute to  inflating the envelope. 

\begin{figure*}
 \centering
\mbox{\includegraphics[width=13cm, angle=-90]{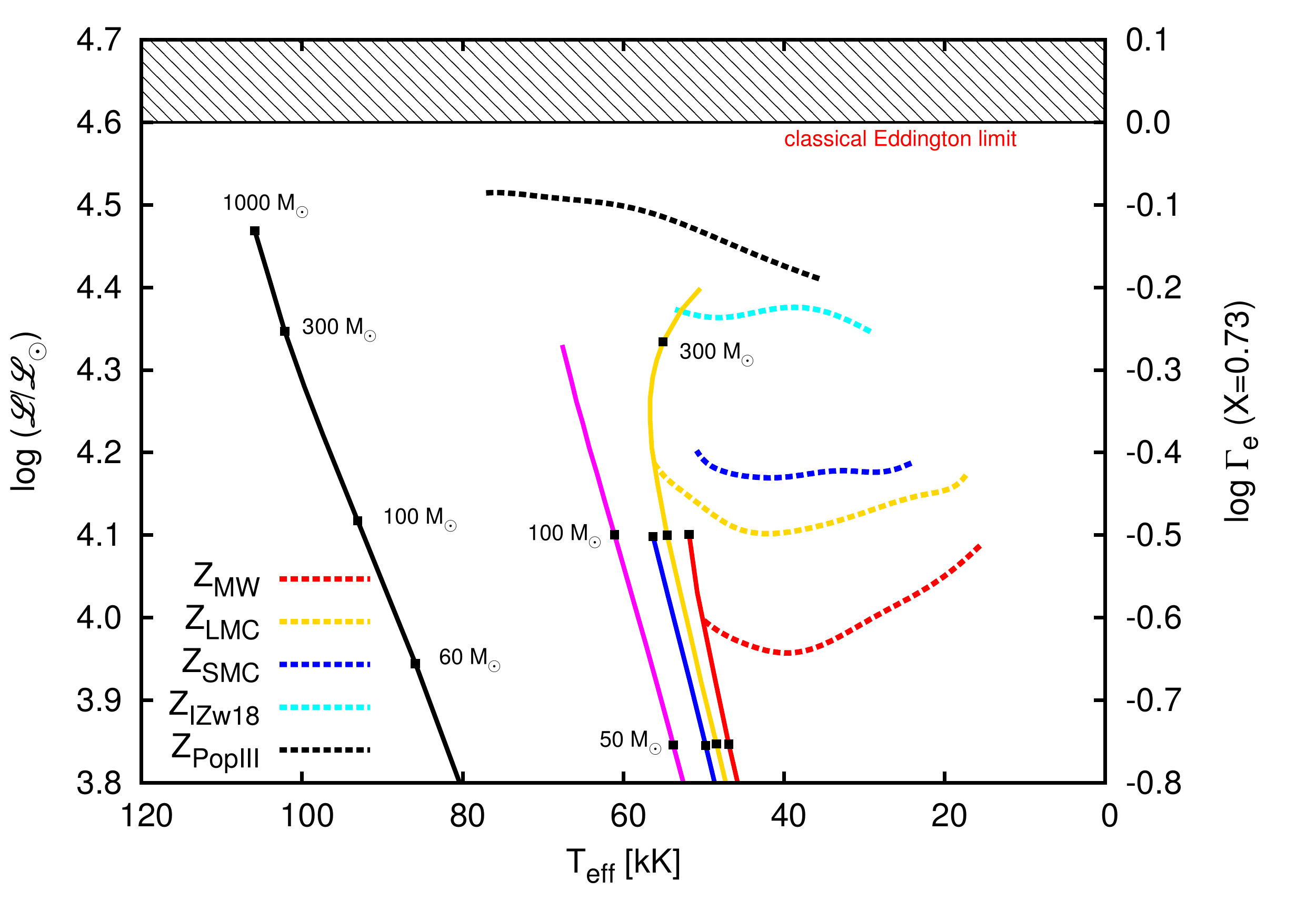}}
   \caption[]{Spectroscopic Hertzprung-Russell diagram showing the boundaries (dotted lines) between non-inflated and inflated models for different metallicities such that below a given line, we do not find any inflated model for that metallicity. The solid lines represent the ZAMS. The masses at ZAMS for  some of the models have been indicated. The right Y-axis represents the logarithm of the classical Eddington factor considering $X=0.73$, similar to Figs.\,\ref{fig:gammax_Z} and \ref{fig:inflation_Z}. The horizontal line marks the location $\Ge=1$ and the hatched region above it is the forbidden zone where no hydrostatic model can lie.}
\label{fig:inflation_lower_bounds}
\end{figure*}

Figure \ref{fig:inflation_lower_bounds} marks the regions in the sHR diagram that separates the non-inflated models from the inflated ones, considering the same sample as in Fig.\,\ref{fig:inflation_Z}. For each model grid the $\teff$ range of the models were divided into 20 equispaced bins and in each bin the un-inflated model with the highest $\mathscr{L}$ was selected. These data points were then joined and the resulting line was smoothed using B\'ezier splines. These lines do not extend to $\teff$ values below $\sim 10\,000\,$K (see Fig.\,\ref{fig:inflation_Z}), because we do not find any core-hydrogen burning model that is not inflated in this temperature range and hence the boundaries cannot be drawn. 

The lines for \zmw and \zlmc show a pronounced dip around $\teff\sim35\,$kK. This is because of the influence of the Fe-bump coupled with inefficient convection, as previously mentioned. At lower temperatures convection becomes more efficient and so these lines move upwards to higher $\mathscr{L}$. For lower metallicities this dip is not clearly identified because the Fe-bump is either weak, or absent.

 In the MW and  LMC grids, the models start to develop inflated envelopes even on the ZAMS, at masses above $\sim80\mso$ and $\sim125\mso$ respectively. At lower metallicities this is also expected to happen, albeit at higher masses and hence at higher $\mathscr{L}$, which is beyond the parameter space explored here.

\subsection{Role of opacity in determining envelope structure}\label{subsec:opacity}

\subsubsection{OPAL opacities}\label{subsubsec:opal}
\begin{figure}
\centering
\mbox{\includegraphics[width=6.5cm,angle=-90]{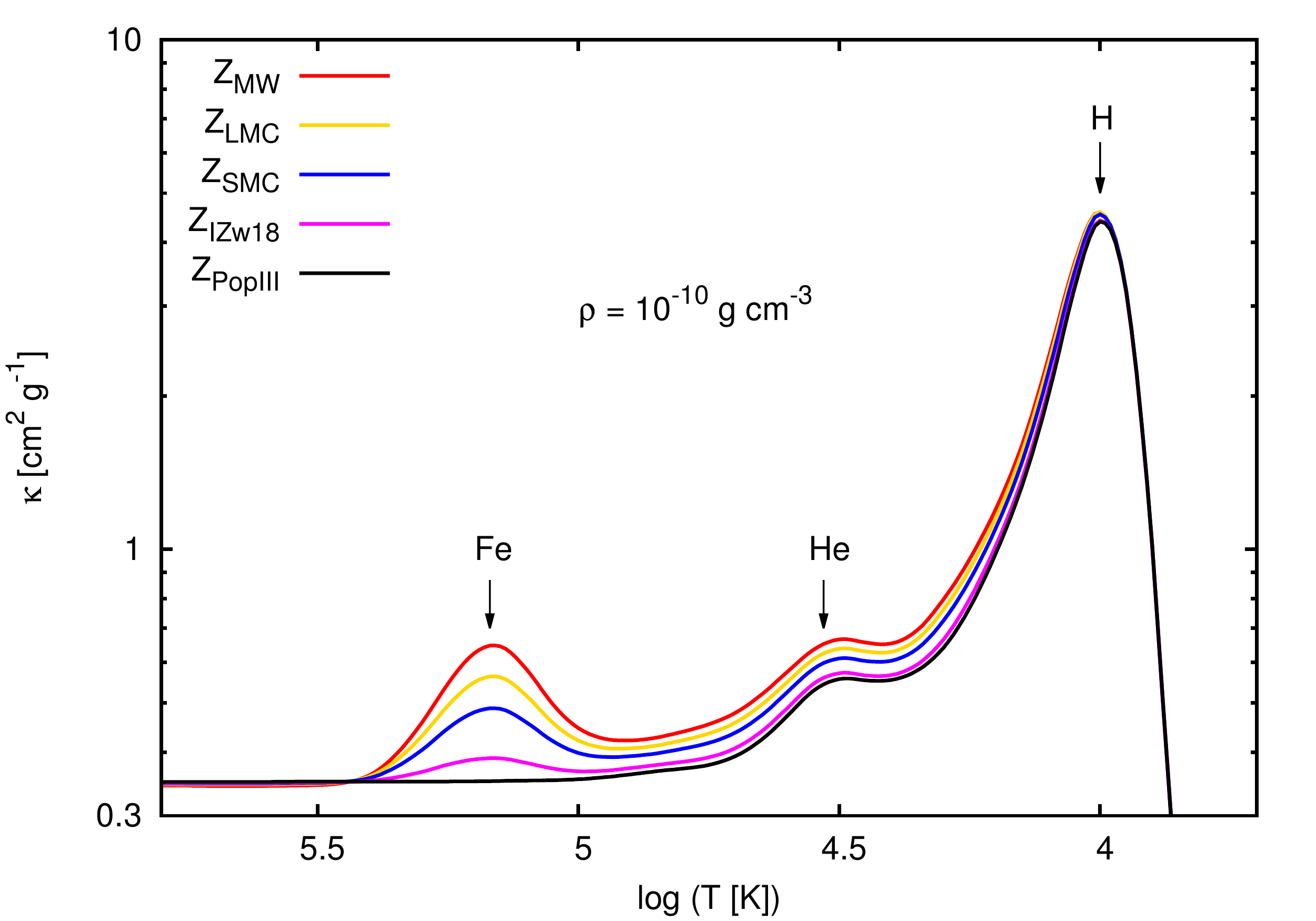}}
   \caption[]{Interpolated opacities from the OPAL tables for $\teff>8000\,$K and from \citet{AF1994} for $\teff>8000\,$K.  The density is fixed at $\rho=10^{-10}\gcm$ and opacities for the five metallicities used in this study are shown. The opacity peaks caused by iron, helium and hydrogen ionisation are marked.}
\label{fig:OPAL_Z_BEC}
\end{figure}

The Rosseland mean opacity $\kappa$ is a function of density, temperature and chemical composition such that for a given $\rho$ and $T$, $\kappa$ increases with an increase in metallicity.  This is demonstrated in Fig.\,\ref{fig:OPAL_Z_BEC} where the three opacity peaks caused by partial ionisation of iron, helium and hydrogen at their characteristic temperatures are visible. Note that the opacity does not vary linearly with metallicity around the Fe bump temperature. The slope $\frac{d\kappa}{dZ}$ is higher for lower values of Z. In this section we  investigate how the strength of these opacity peaks determine the density structure of the inflated envelope. 

\begin{figure}
\centering
\mbox{\includegraphics[width=6.5cm,angle=-90]{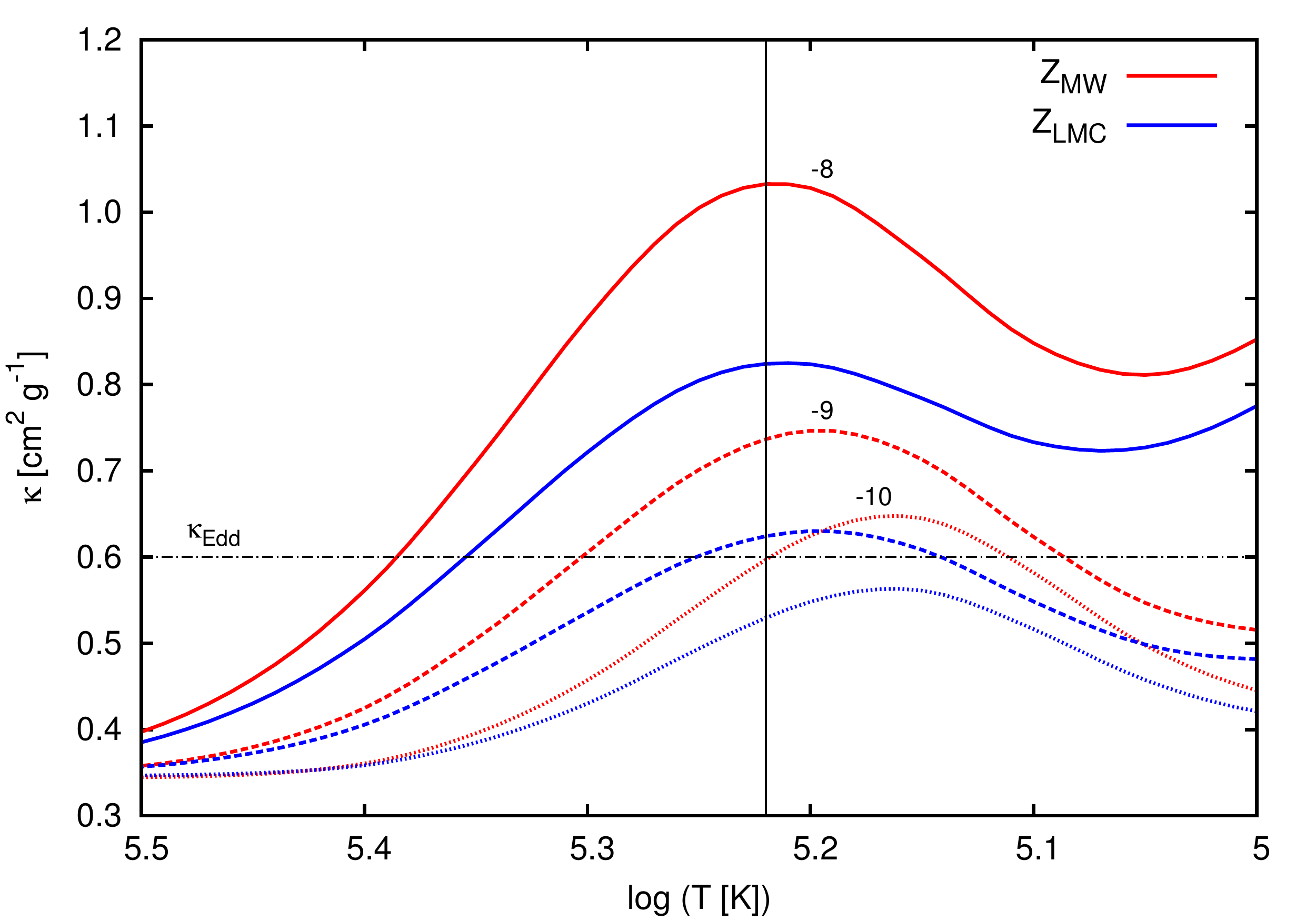}}
   \caption[]{OPAL opacities for two metallicities \zmw and \zlmc, and three different values of $\log (\rho[\rm g\,cm^{-3}])$ each, i.e., $-8$, $-9$ and $-10$, as indicated in the plot. The dot-dashed line at $\kappa=0.6$ is the assumed location of $\kappa_{\rm Edd}$.}
\label{fig:OPAL_Z}
\end{figure}

In Fig.\,\ref{fig:OPAL_Z}, we take a look at the OPAL opacities around the Fe-bump for the MW and LMC metallicities. As mentioned  before, in the inflated envelope the condition $\Gamma\approx 1$ holds true. Let the corresponding opacity be $\kappa_{\rm Edd}$ such that $\Gamma = \kappa_{\rm Edd}L_{\rm rad}/4\pi cGm \approx 1$. Consider two models with the same $L/M$ but with metallicities \zmw and \zlmc such that $\kappa_{\rm Edd}=0.6$ (dot-dashed line), and assume that the convective efficiency is negligible. At the peak of the Fe-bump marked by the vertical black line, the MW model has to decrease its density by two orders of magnitude, from $10^{-8}\gcm$ to $\approx10^{-10}\gcm$ whereas the LMC model only has to go down to $\sim 10^{-9}\gcm$ to satisfy the constraint $\Gamma =1$. Therefore, the higher metallicity model will  adjust its envelope structure such that it has a lower envelope density. In practice however, convection may mediate this effect \citep{sanyal2015}.

\subsubsection{Opacity in the inflated envelope}\label{subsubsec:opac_infl}

 \begin{figure}
\centering
\mbox{\includegraphics[height=9.4cm,  width=14cm, angle=-90]{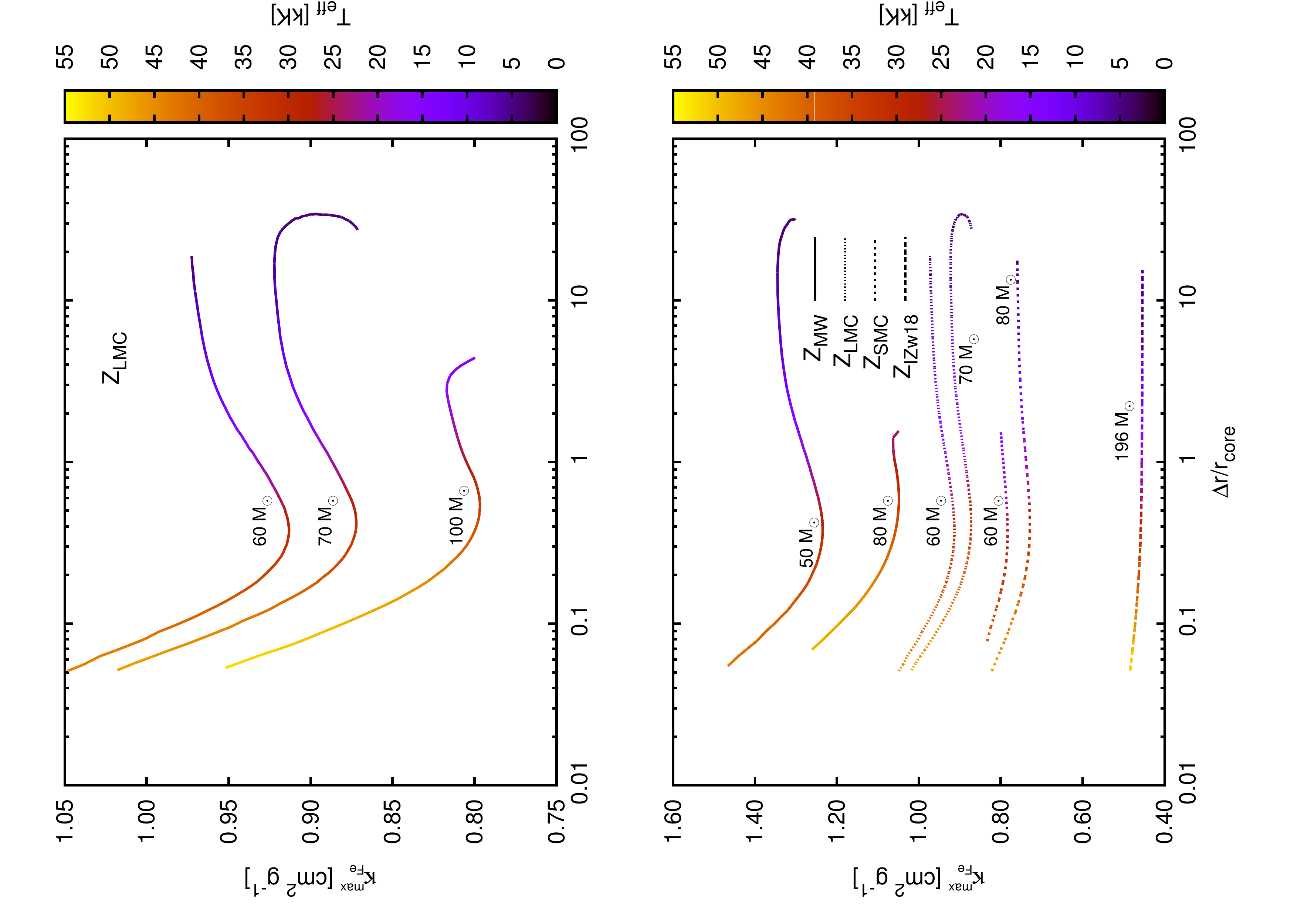}}
   \caption[]{{\it Top:} Maximum opacity within the Fe-bump region for three different evolutionary sequences with initial masses $60,70$ and $100\mso$ from the LMC grid, as a function of \inflation. The effective temperature of the model is colour-coded. Only the part of the evolution where $Y_{\rm s}<0.3$ has been plotted. 
   {\it Bottom:} Same as in the top panel, but for four metallicities.}
\label{fig:kappa_T_Z}
\end{figure}

As mentioned before, the opacity bumps caused by the partial ionisation zones at characteristic temperatures play a major role in determining the structure of an inflated envelope. As prototypical examples, we have selected three sequences with $Z=Z_{\rm LMC}$ and with initial masses of  $60\mso,70\mso$ and $100\mso$. The base of the inflated envelope in these models is located around the characteristic Fe-bump temperature  $T_{\rm Fe}\approx170\,000$ K. The maximum opacity within the Fe-bump (\kmax), i.e. between $5<\log (T/K)<5.5$, for the three sequences are shown in the top panel of Fig.\,\ref{fig:kappa_T_Z}, for that part of the evolution where the models are not helium-enriched at the surface, i.e. $Y_{\rm s}<0.3$. At any given value of \inflation, the higher mass model has a lower \kmax because it has higher luminosity, and hence needs to decrease its opacity further to maintain $\kappa\approx\kappa_{\rm Edd}$.

The $60\mso$ sequence for example develops a larger inflated envelope as it evolves, while increasing its $\mathscr{L}$. The opacity within the Fe-bump and \kmax therefore decrease in the initial phase because convection is relatively inefficient. As the model evolves to cooler effective temperatures, the Fe-bump goes deeper inside the star where densities are higher, and convection becomes efficient. Hence, \kmax increases at $\teff\lesssim25\,000\,$K. In the case of the $70\mso$ sequence however, there is a drop in \kmax at $\teff<5000\,$K. In this phase of the evolution, a high mass-loss rate ($\sim10^{-5}\msoy$) causes a sharp increase in $\mathscr{L}$. As a result, $L_{\rm rad}$ increases in the Fe-bump region. But the convective efficiency does not increase enough (for details, see Appendix \ref{app:70mso}) such that it can prevent $\kappa_{\rm Fe}$ from  going down.

Models with a higher metallicity have a stronger Fe-bump, the effect of which is seen in the bottom panel of Fig.\,\ref{fig:kappa_T_Z}. While $\kmax$ for the $80\mso$ $Z_{\rm MW}$ model with the highest $\teff$ is $\sim1.2\,{\rm cm^2\,g^{-1}}$, the same quantity for the $196\mso$ $Z_{\rm I\,Zw18}$ model is $\sim 0.5\,{\rm cm^2\,g^{-1}}$. The slope of $\kappa_{\rm Fe}^{\rm max}$ versus \inflation\, is steeper for the \zmw evolutionary sequences compared to the other sequences at lower metallicities because of the nature of the OPAL opacities explained in Sec.\,\ref{subsubsec:opal}.

\subsection{Mass contained in the inflated envelopes}

 \begin{figure}
\centering
\mbox{\includegraphics[width=6.5cm, angle=-90]{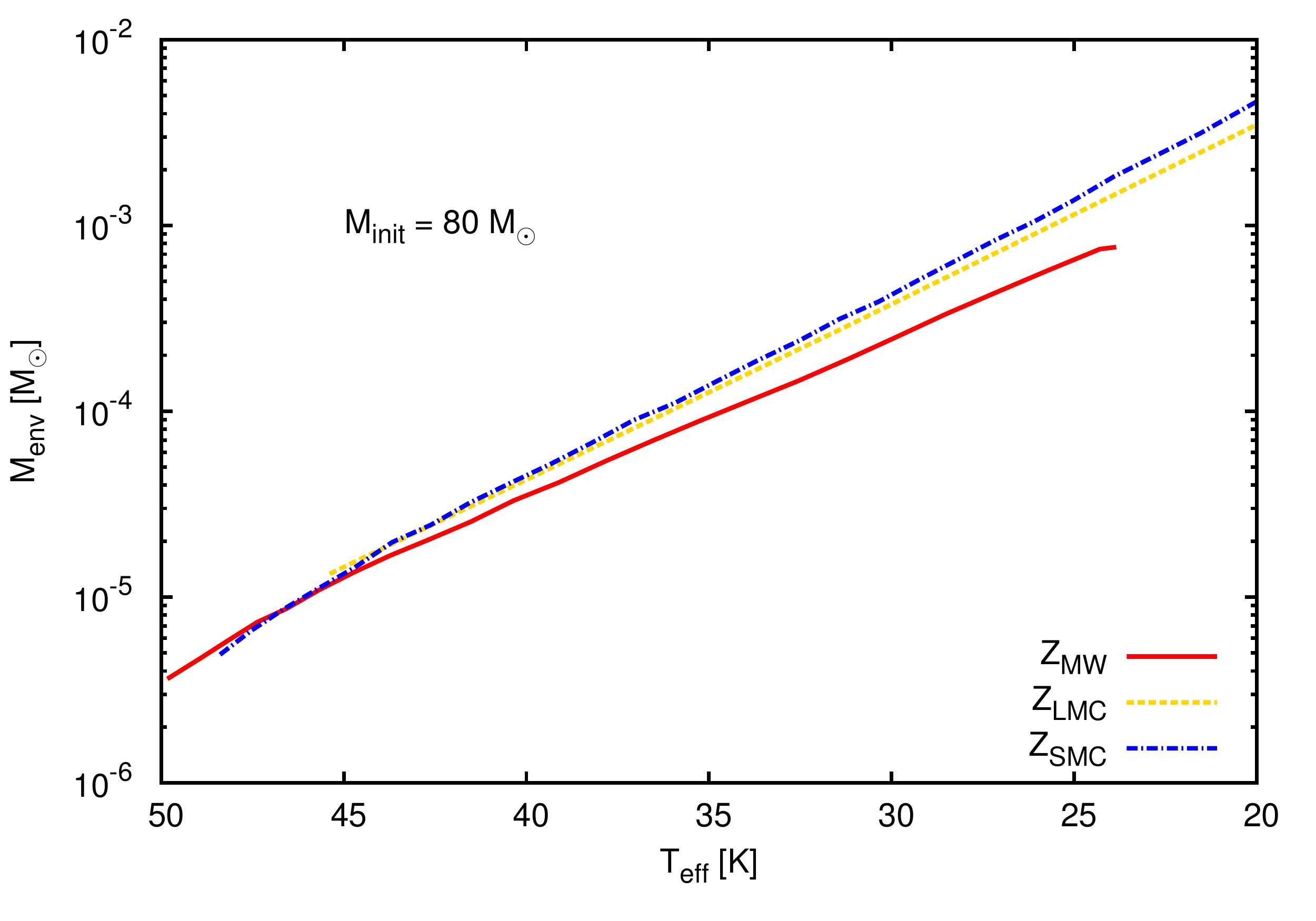}}
   \caption[]{Envelope mass versus effective temperature for the $80\mso$ sequences from the MW, LMC and SMC grids. Only models in the $\teff$ range $20-50\,$kK and with $Y_{\rm s}<0.3$ are shown. }
\label{fig:envmass_teff}
\end{figure}

In this section we investigate the inflated envelope masses of our models.  In Fig.\,\ref{fig:envmass_teff} we compare the $80\mso$ sequences in the MW, LMC and SMC model grids and show that for a given $\teff$, the higher metallicity model  has a lower envelope mass. At relatively high effective temperatures ($\teff>45\,000\,$K), i.e. when the sequences start developing inflated envelopes for the first time during their evolution, the envelope masses for all three sequences are similar but at as they evolve to lower $\teff$, the distinction becomes clear. For example, at $\teff=30\,000\,$K the $80\mso$ MW model has a distinctly smaller envelope mass  than the corresponding LMC and SMC models, the difference  in their core radii being negligible.  The LMC and the SMC models however have comparable envelope masses over the whole $\teff$ range.  This  trend is likely related to the relative strength of the iron opacity peaks for these metallicities (cf.\,Fig.\,\ref{fig:OPAL_Z_BEC}).

 \begin{figure}
\centering
\mbox{\includegraphics[width=14cm,height=10.5cm, angle=-90]{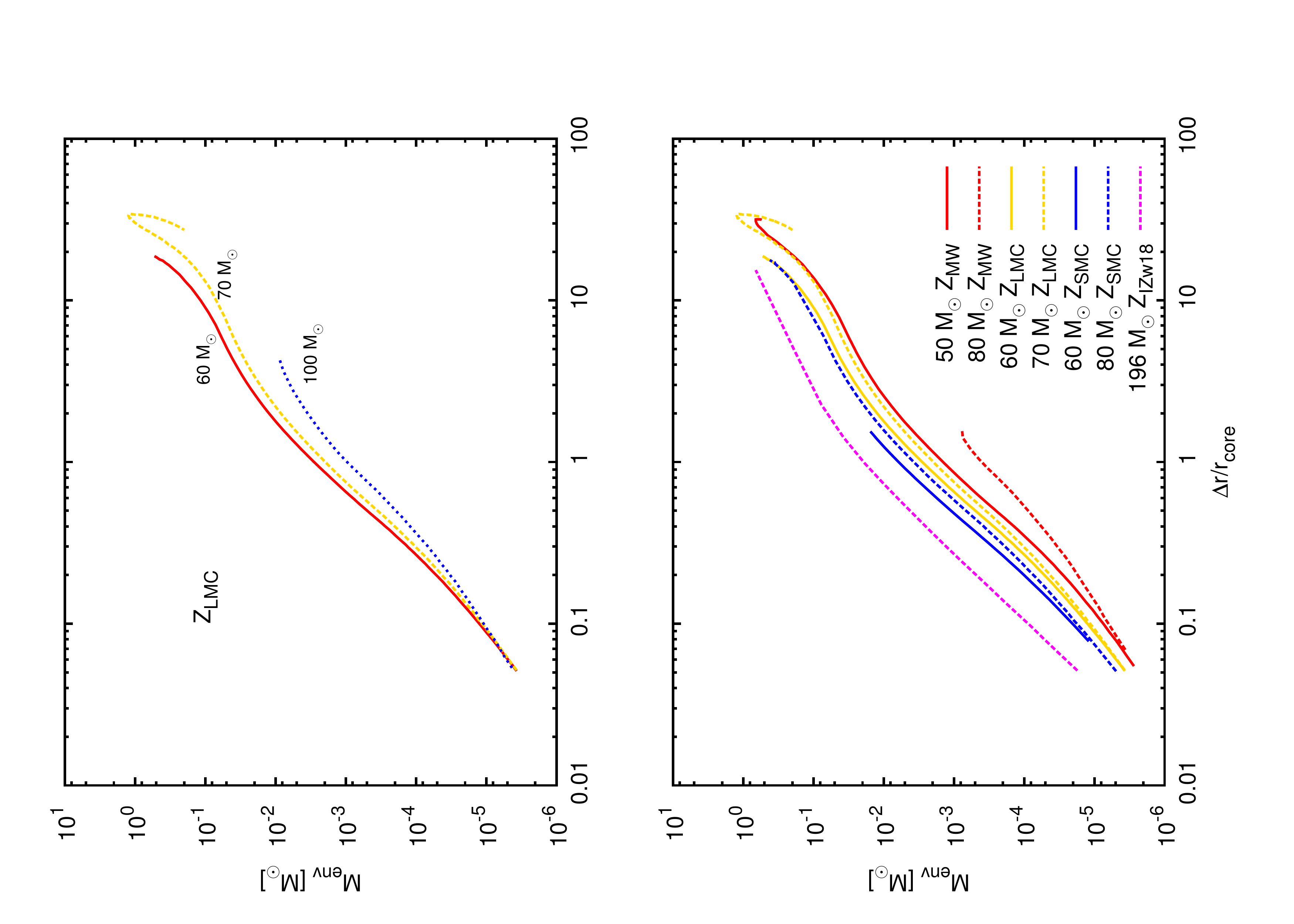}}
   \caption[]{{\it Top:} Inflated envelope masses for 3 sequences from the LMC grid as a function of \inflation, such that $Y_{\rm s}<0.3$. \\
   {\it Bottom:}  Same as in the top panel, but for four metallicities.}
\label{fig:envmass_Z}
\end{figure}

For evolutionary sequences of a given metallicity, say, $Z_{\rm LMC}$,  the ones with higher $\mathscr{L}$'s have lower envelope masses ($M_{\rm env}$), as shown in the top panel of Fig.\,\ref{fig:envmass_Z}. For small inflation, i.e. \inflation$<0.1$, the envelope mass in the three LMC   sequences is comparable but as \inflation\, increases, the sequences separate out such that for a given \inflation, the $100\mso$ sequence with the highest $\mathscr{L}$ has the lowest envelope mass and comparing with Fig.\,\ref{fig:kappa_T_Z}, the lowest $\kappa_{\rm Fe}^{\rm max}$. We note that for the $70\mso$ sequence for example, the envelope mass varies by more than five orders of magnitude over its main-sequence lifetime. The drop in $M_{\rm env}$ near the end of the $70\mso$ sequence is because of its blueward evolution in the HR diagram caused by strong mass-loss (cf.\,Sec.\,\ref{subsec:inflation_Z}).

 \begin{figure}
\centering
\mbox{\includegraphics[width=14cm,height=10.5cm, angle=-90]{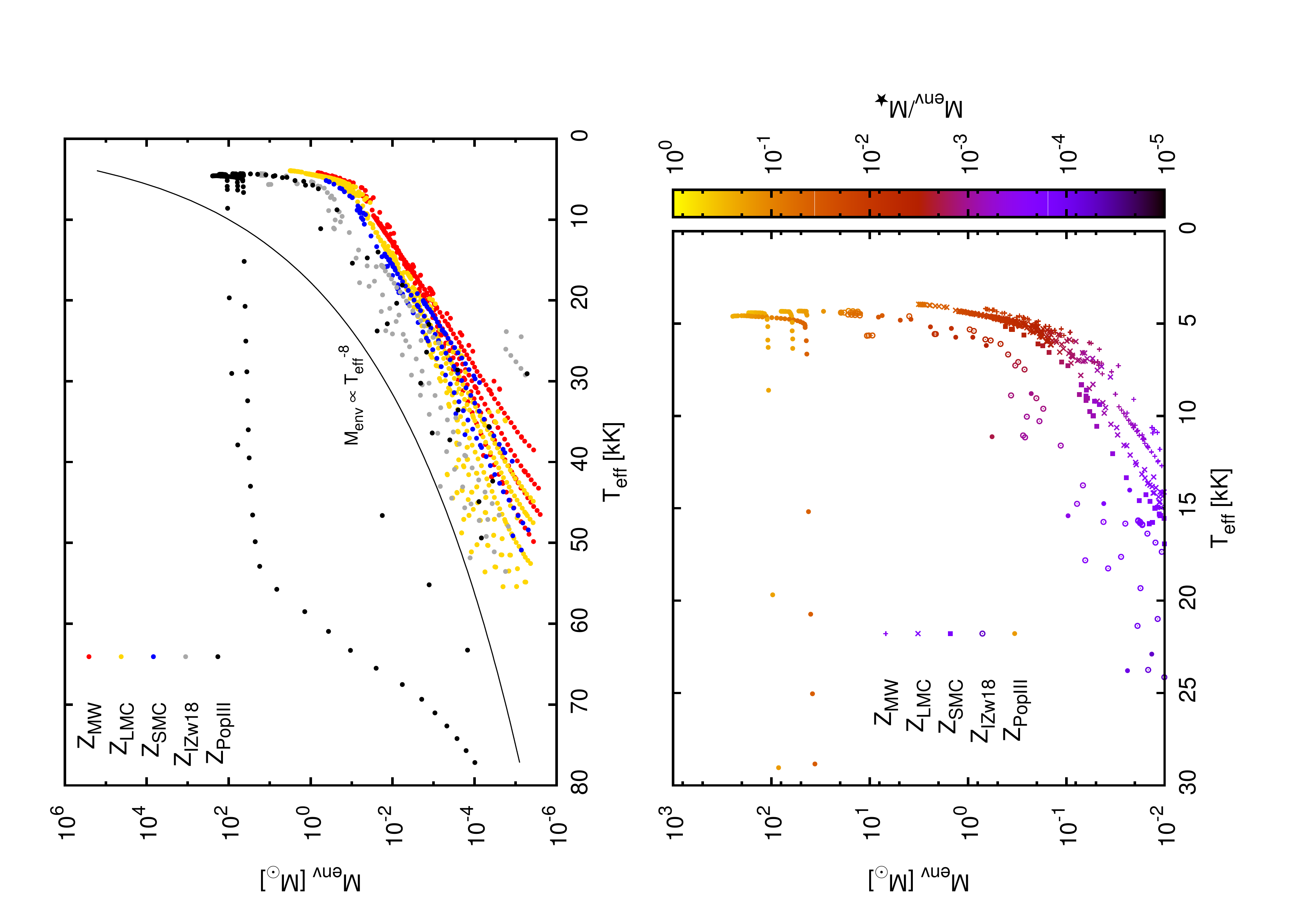}}
 \caption[]{Mass contained in the envelope of inflated models as a function of their effective temperatures, for the five  metallicities considered in this study. Only models with $Y_{\rm s}<0.3$ are shown. The black dotted line has been drawn for guiding the eye (see text). The bottom panel only shows models with $\menv>10^{-2}\mso$ and $\teff<30\,000\,$K. The colour bar on the right represents the fraction of the stellar mass contained in the inflated envelope.}
\label{fig:envmass}
\end{figure}

In the bottom panel of Fig.\,\ref{fig:envmass_Z}, several representative sequences from the  MW, LMC, SMC and \izw grids are shown that depicts how $M_{\rm env}$ changes with inflation. The higher metallicity models have lower envelope mass for a fixed \inflation. For example, at $\inflation\,=1$, the $60\mso$ LMC sequence has $M_{\rm env}=3\times10^{-3}\mso$ while the $60\mso$ SMC sequence has $M_{\rm env}=6\times10^{-3}\mso$. At high inflation ($\inflation>10$), some of the lines touch each other which may be related to the different $L/M$-ratios of the models induced by mass-loss.

We investigate the envelope masses of all the inflated models in  Fig.\,\ref{fig:envmass}. The envelope  mass spans several orders of magnitude from $\sim 10^{-5}\mso$ to $\sim 100\mso$. In general we find that \menv increases with a decrease in $\teff$ for a given metallicity. This increase in \menv is distinctly steeper at $\teff\lesssim8000\,$K (bottom panel of Fig.\,\ref{fig:envmass}) compared to that at $\teff>10\,000\,$K.   Below $10\,000$ K, the low-$Z$ models have very massive envelopes (bottom panel of Fig.\,\ref{fig:envmass}). The models which contain the hydrogen opacity bump show strong density inversions \citep{sanyal2015} and because of the sharp rise in density the envelope mass increases.

At $\teff>10\,000\,$K, there is a spread in $M_{\rm env}$ over a few orders of magnitude, but at the lowest effective temperatures the spread is much narrower. This is because the sequences which evolve to effective temperatures below $\sim 8000\,$K do so for a narrow mass range. At higher initial masses strong mass-loss prevents them from evolving to low surface temperatures, and at lower initial masses inflation is not strong enough.

\citet{graefener_2012} found from analytical estimates that the inflated envelope mass scales as $M_{\rm env}\sim R^4/M$. For constant $M$ and $L$, this translates to $M_{\rm env}\sim \teff^{-8}$.  The shape of this curve (Fig.\,\ref{fig:envmass}) is well-reproduced by our model grids at higher $\teff$, although at $\teff<8000\,$K the dependence is steeper than the analytical estimate. We note that the envelope mass estimates of the $M_{\rm init}=1000\mso$ and the $M_{\rm init}=500\mso$ models in the \pop3 grid (some of these are the black dots located above the dotted line in Fig.\,\ref{fig:envmass}) are particularly uncertain because they are sensitive to the choice of the threshold value of $\beta$ that marks the location of $r_{\rm core}$. Although these models have inflated envelopes, the absence of an opacity peak complicates the process of identifying $r_{\rm core}$.   A more detailed investigation of the envelope structures of massive \pop3 models is beyond the scope of the present paper and will be pursued in a forthcoming study. 

The envelope mass is determined both by the extent of inflation ($\Delta r/r_{\rm core}$) and the metallicity. As the metallicity increases, the models with the most massive inflated envelopes (the cool supergiants) are found at lower masses. This is a consequence of the applied mass-loss rates. The wind mass-loss prescriptions used in stellar evolution calculations are functions of luminosity, temperature, mass, radius and chemical composition of the model. With an increase in luminosity or mass, the wind mass loss rates increase and the most massive stars in our MW and LMC grids become helium-rich WR stars  \citep{koehler2015} and do not become cool enough to contain massive envelopes ($\gtrsim1\mso$), as explained in the previous paragraph. At lower $Z$ this happens at higher masses. At $Z=0$, models never become helium-rich at the surface unless they are very fast rotators \citep{yoon2012}.


\section{Discussion and conclusions}\label{sec:conclusions}

We have performed a study of the envelope structures of core-hydrogen burning massive star models computed with the following metallicities: \zmw, \zlmc, \zsmc, \zizw and \zpop3. We investigated the Eddington factors in their interior and its connection to envelope inflation as a function of metallicity.

As expected we found that the Eddington limit is metallicity dependent such that models with a higher $Z$ reach $\Gamma=1$ in their interior at a lower mass. While a $30\mso$ MW model reaches $\Gamma\approx1$ in its interior, it requires a  $150\mso$ Pop\,III model to obtain similar Eddington factors on the hot side of the HR diagram, i.e., at $\teff>10\,000\,$K. For models with $\teff$ below the hydrogen recombination temperature, metallicity has little effect, and super-Eddington layers can be found down to $\sim 5\mso$ models, although in the post main-sequence phase \citep{langer2015_proc,grassitelli2015b}. Proximity to the Eddington limit leads to envelope inflation in our models. We find inflated models at all the metallicities investigated, albeit at different $L/M$-ratios (Fig.\,\ref{fig:inflation_lower_bounds}).  At a higher $Z$, envelope inflation starts at lower masses because of larger opacities that help approach $\Gamma\approx1$.  We reiterate that envelope inflation might already start to develop before reaching the Eddington limit because of the contribution from the gas pressure gradient (cf.\,Sec.\,\ref{sec:inflation}).   Envelope  inflation is responsible for the redward bending of the ZAMS and the TAMS in the upper HR diagram (Fig.\,\ref{fig:ZAMS}), that is also supported by observations which show how the upper part of the Galactic H-R diagram is well-populated by stars up to $\teff\sim10\,000\,$K \citep{castro_2014}. The extent of inflation might be used to infer the value of $\alpha_{\rm MLT}$ for massive stars by comparing the main-sequence width of the models against the observational TAMS \citep{castro_2014,bestenlehner2014}.

We find that the mass contained in the inflated envelopes can range from $\sim10^{-6}\mso$ in the hot, luminous models to $\sim 100\mso$ in the cool supergiant type models, across the range of metallicities investigated. While the observational signatures of these envelopes needs to be explored further, the ones with high envelope masses ($M_{\rm env}>1\mso$) seem to be promising candidates for explaining the violent LBV eruptions, for e.g., the 1860 outburst $\eta$ Car, and other  $\eta$ Car analogs  \citep{khan2015} or supernova imposters. These models are near the Eddington limit and have several solar masses in the loosely bound envelope. The details of the instability responsible for the outburst still needs to be investigated.  On the other hand, if the inflated envelopes are lost episodically from the models with small envelope masses, it will cause them to shrink to the non-inflated core radius but will not be able to change the bolometric luminosity appreciably. These models have been put forward to explain  the S-Doradus type variations by \citet{graefener_2012} and \citet{sanyal2015}.

\citet{moriya2015} proposed that an observational consequence of a supernova progenitor with an inflated envelope is that it extends the rise time of the supernova shock-breakout signal. This naturally explains the long ($\sim50\,$s) shock breakout X-ray signal detected from the Type Ic SN\,2008D \citep{soderberg2008} that is believed to have had a compact WR progenitor. 

Luminous helium stars also show pronounced core-halo structures and such models have been investigated in the past \citep{ishii99,petrovic_2006,graefener_2012,tramper2015,grassitelli2016a}. The apparent mismatch in radii between model atmosphere calculations and stellar interior models of massive Galactic Wolf-Rayet stars  has been claimed to have been reconciled by envelope inflation \citep{graefener_2012}.

The inflated models are potentially unstable against the so-called strange-mode instability \citep{gautschy&glatzel90,glatzel&kiriakidis_93b} because of low heat capacities in their dilute envelopes \citep{glatzel1994}. \citet{glatzel&kiriakidis_93b} reported that their solar metallicity models with $\log(\shrL)\gtrsim 4$ are unstable to strange-mode oscillations. This result coincides with the boundary between the inflated and non-inflated models in our \zmw grid. Furthermore, these oscillations might drive mass-loss from the star \citep{grott2005} though \citet{moriya_langer_2015} and \citet{grassitelli2016a} find mass-loss to dampen the pulsations.  The pulsational properties of our models will be explored in detail in a forthcoming study.

A critical ingredient in the physics of envelope inflation is convection, i.e. how convective energy transport is treated in these regions. In the literature, stellar models computed with increased convective efficiency show little or no envelope inflation \citep{ekstroem2012,yusof2013}.  A discussion of the convective efficiencies in our \zlmc models can be found in Sec.\,6 of \citet{sanyal2015}.  \citet{jiang2015} performed 3-D radiation hydrodynamics simulations of massive star envelopes and concluded that for a $80\mso$ ZAMS model, standard MLT overestimates the convective flux in the inflated region around the Fe-bump. In that case inflation in our 1-D models has been underestimated. \citet{jiang2015} also found turbulent velocities that exceed the isothermal sound speed, driving shocks in the envelope and creating an inhomogeneous, clumpy medium which, however, do not lead to a break-down of the inflation.

\citet{grassitelli2016b} recently investigated the role of turbulent pressure ($P_{\rm turb}$) in stellar models computed with MW, LMC and SMC metallicities, and found that its effect on stellar structure is negligible regardless of the metallicity \citep{grassitelli2015}.  However, the ratio  of $P_{\rm turb}$  to $P_{\rm total}$ in the stellar envelopes of the hot stellar models decreases for lower metallicities at a given temperature and luminosity. This trend is consistent with our results for inflation (Figs.\,\ref{fig:inflation_Z} and \ref{fig:inflation_lower_bounds}). At higher metallicities the density in the inflated envelope is lower  which implies inefficient convection and therefore a large and negative entropy gradient. Hence the convective velocities and the Mach number is also higher which leads to higher turbulent pressure.

Furthermore, \citet{grassitelli2015}  found a correlation between macroturbulent velocities in Galactic OB stars and the fraction of turbulent pressure in the stellar envelope models. Since the turbulent pressure contribution in the inflated envelope becomes stronger in the upper HR diagram, high macroturbulent velocities \citep[$\gtrsim 50 \kms$,][]{sergio2015,grassitelli2015,sergio2016} might well be a signature of envelope inflation in hot, massive stars. The conditions in the inflated envelope might be inferred via asteroseismic studies \citep{aerts2014}, especially if the connection between inefficient convection and high-order non-radial pulsations is confirmed \citep{aerts2009,grassitelli2015b,grassitelli2015}.

It might be interesting to look at the fate of the inflated envelopes in close binaries, since $\sim 70$\% of all massive stars are believed to interact during their lifetimes \citep{sana_2012}. The loosely bound envelopes might help to stabilise mass-transfer in close massive binary systems, especially in metal-rich systems where this is expected to happen at lower masses. In close binaries, the hydrogen envelope is usually lost from the mass donor that bares its helium core and increases the $L/M$ ratio. Helium stars with solar metallicity start to develop inflated envelopes from $\sim 10 \mso$ (see Fig.\,19 in \citet{koehler2015}). Massive Type Ib/c progenitors in binary systems are thus expected to have inflated envelopes \citep{yoon2010}.

\begin{acknowledgements}
D. Sz\'ecsi was supported by GA\v{C}R grant 14-02385S. SCY acknowledges support from the Korea Astronomy and Space Science Institute under the R\&D program (Project No. 3348- 20160002) supervised by the Ministry of Science, ICT and Future Planning.
\end{acknowledgements}

\bibliographystyle{aa}
\bibliography{debashis_refs}


\clearpage
\appendix
\section{Evolution of a $70\mso$ inflated LMC model}\label{app:70mso}
We present the evolution of a typical inflated model, the $70\mso$ LMC sequence, with respect to its inflated envelope and the properties around the iron opacity bump. Only that part of the evolution has been studied where the surface helium mass fraction ($Y_{\rm s}$) is lower than $0.3$.

\begin{figure}
\centering
\mbox{\includegraphics[width=6.5cm,angle=-90]{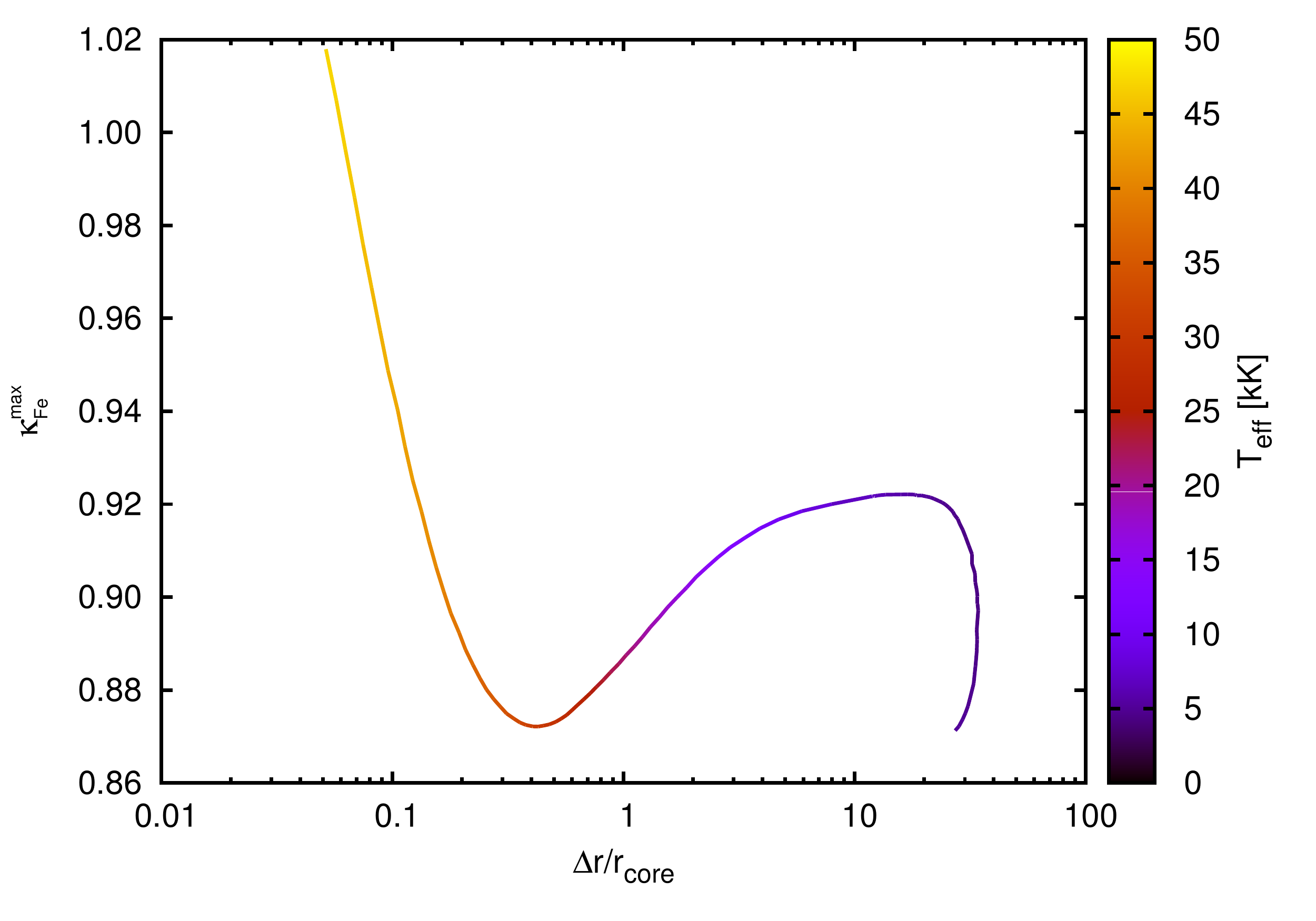}}
   \caption{Maximum opacity in the temperature range $5<\log (T/K)<5.5$ as a function of \inflation for that part of the evolution where $Y_{\rm s}<0.3$. The colour bar indicates the effective temperature of the models.}
\label{appfig:kappa}
\end{figure}

\begin{figure}
\centering
\mbox{\includegraphics[width=6.5cm,angle=-90]{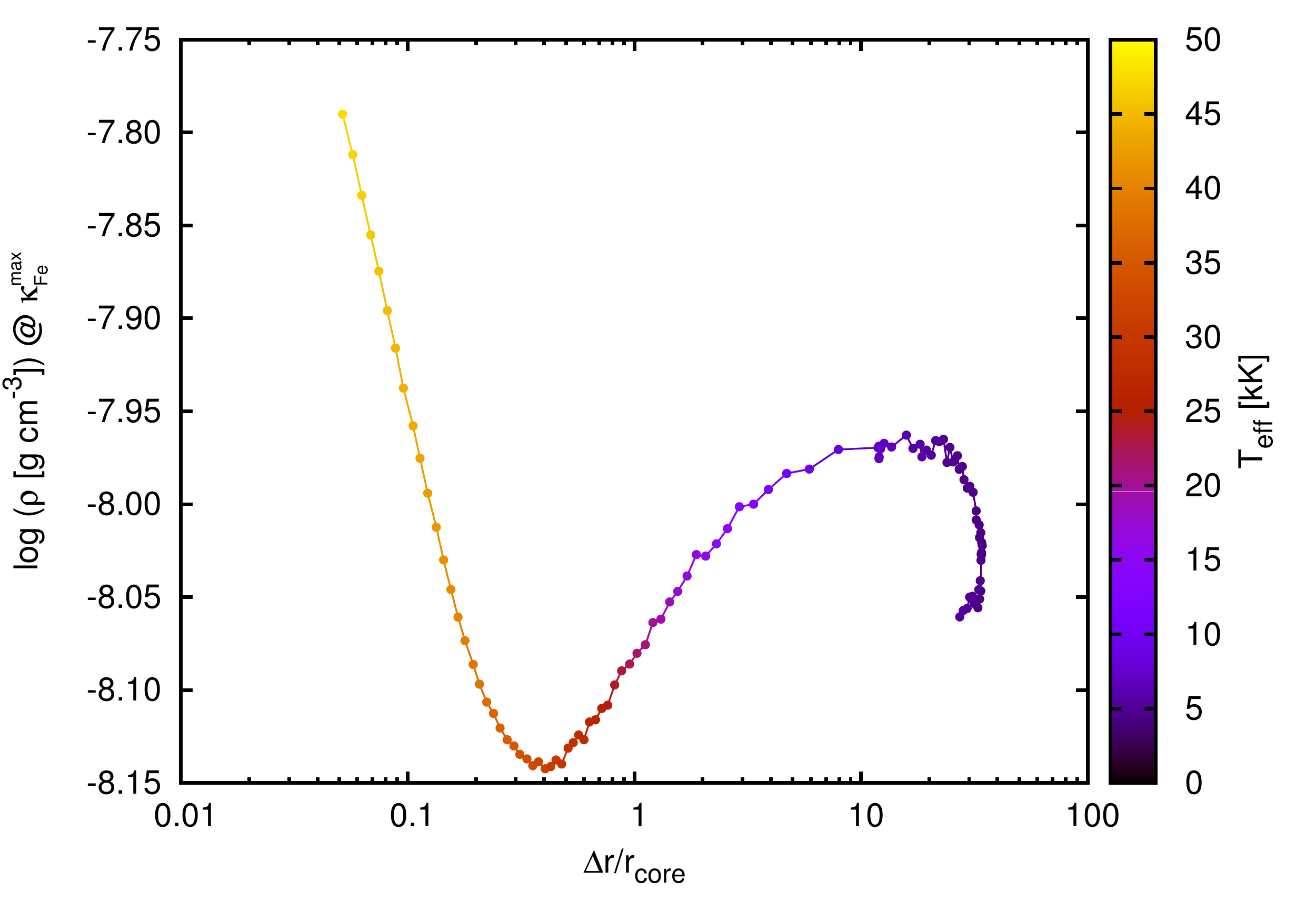}}
   \caption{The variation of density at the position of $\kmax$ as a function of \inflation, for that part of the evolution where $Y_{\rm s}<0.3$. The colour bar indicates the effective temperature of the models.}
\label{appfig:dens}
\end{figure}

\begin{figure}
\centering
\mbox{\includegraphics[width=6.5cm,angle=-90]{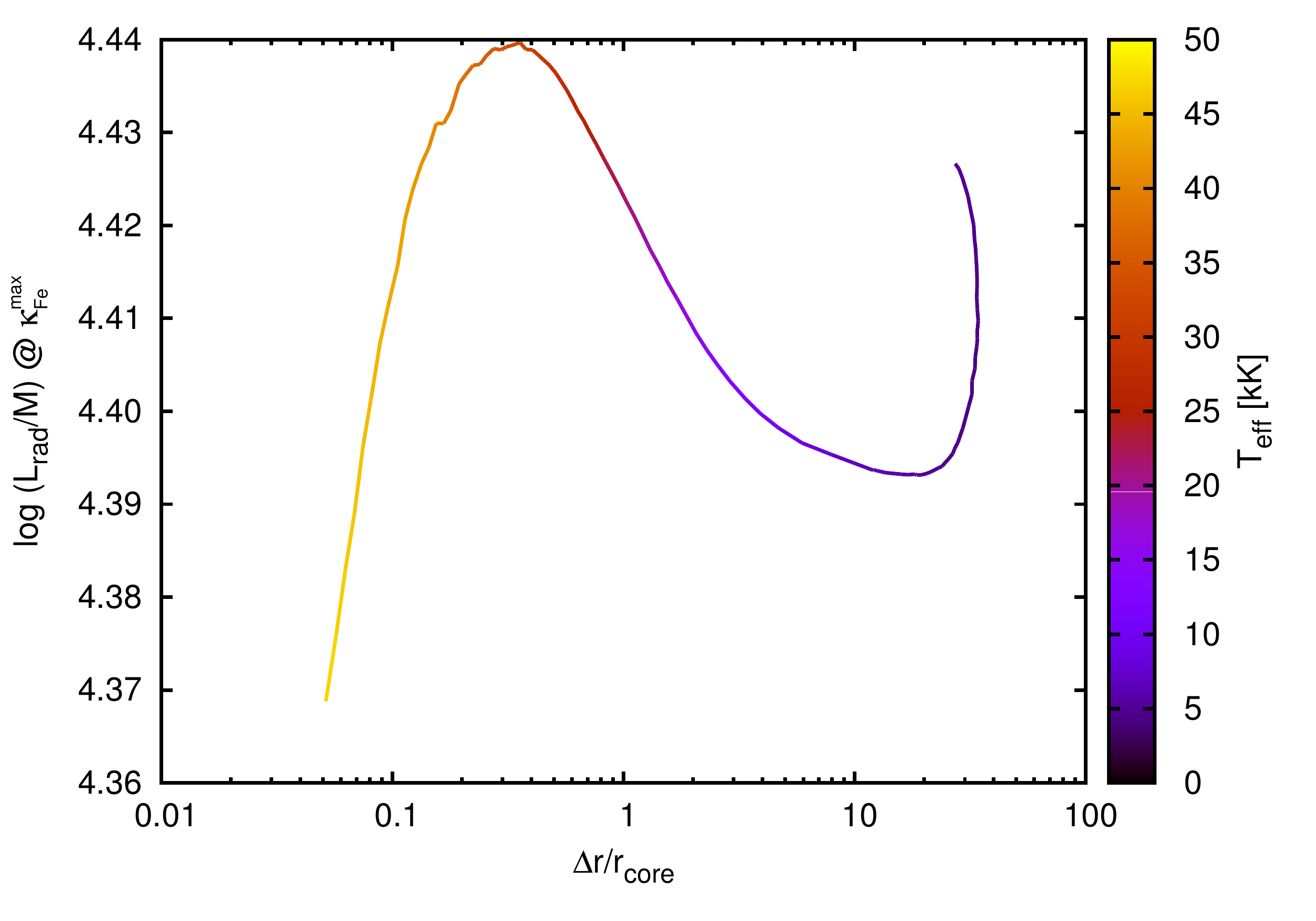}}
   \caption{The $L_{\rm rad}/M$-ratio at the position of $\kmax$  as a function of \inflation, for that part of the evolution where $Y_{\rm s}<0.3$. The colour bar indicates the effective temperature of the models.}
\label{appfig:Lrad}
\end{figure}

\begin{figure}
\centering
\mbox{\includegraphics[width=6.5cm,angle=-90]{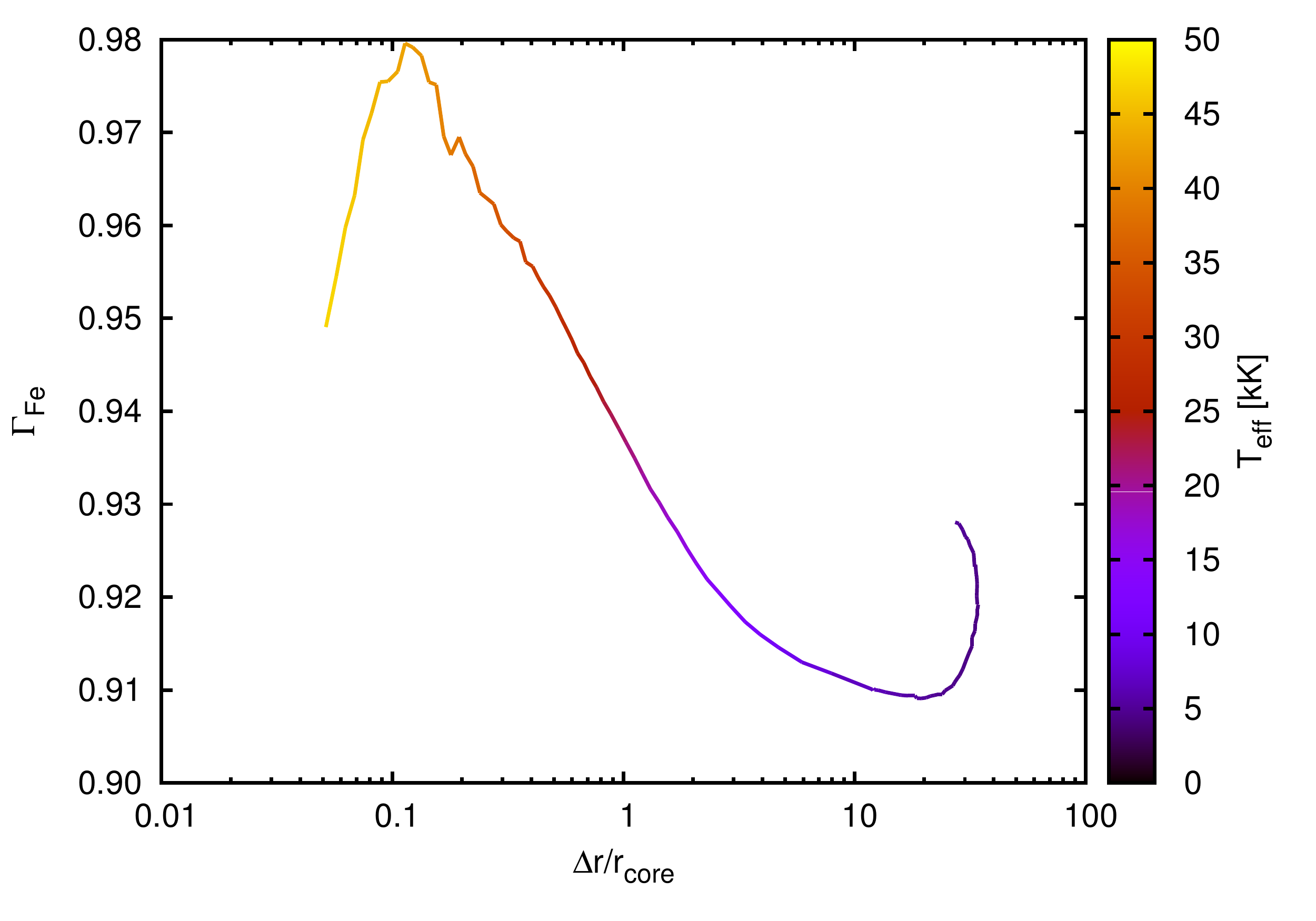}}
   \caption{The value of the Eddington factor at the position of $\kmax$ (denoted as $\Gamma_{\rm Fe}$) as a function of \inflation, for that part of the evolution where $Y_{\rm s}<0.3$. The colour bar indicates the effective temperature of the models.}
\label{appfig:gamma}
\end{figure}

\begin{figure}
\centering
\mbox{\includegraphics[width=6.5cm,angle=-90]{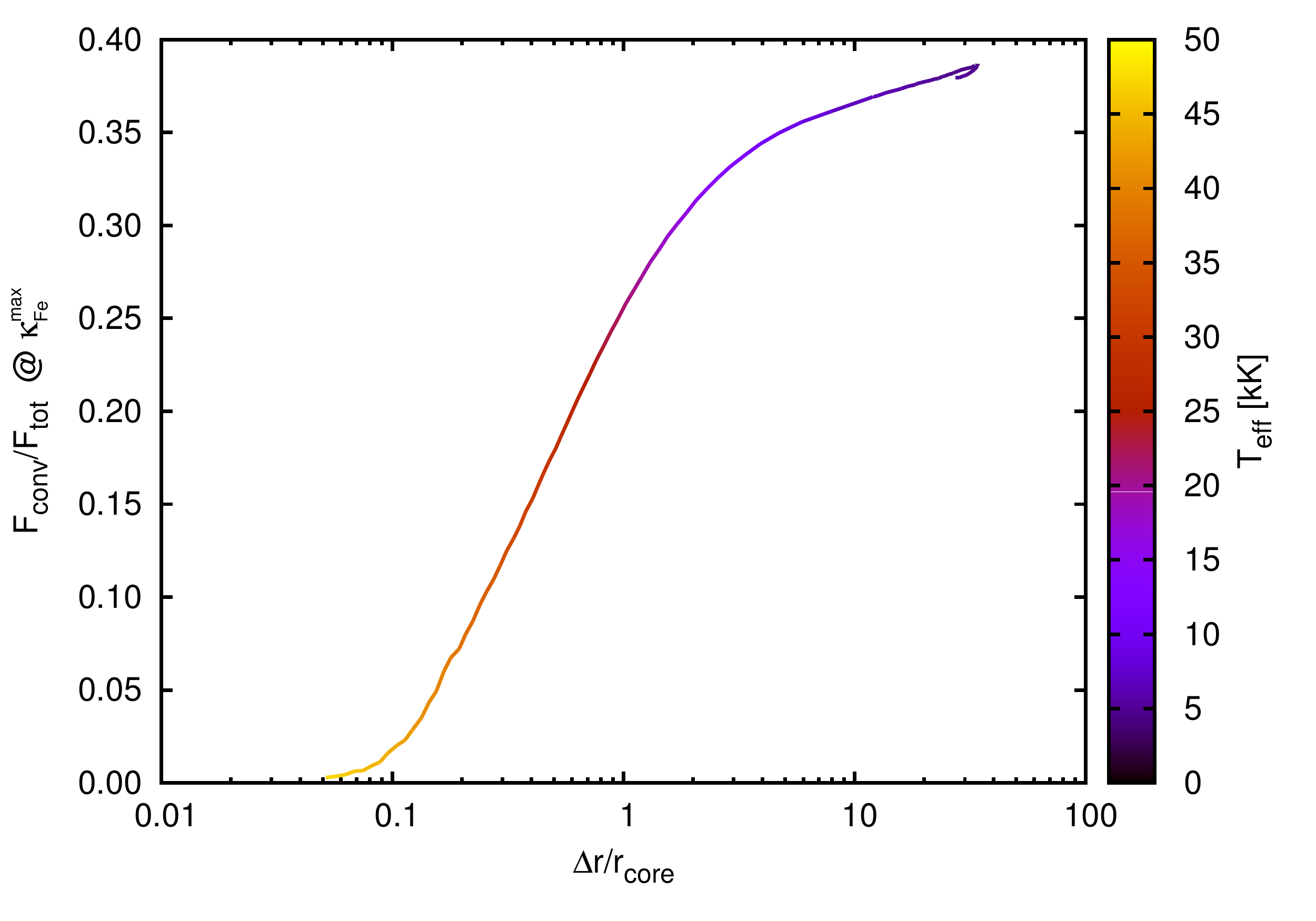}}
   \caption{The ratio of convective flux to the total flux  at the position of $\kmax$  as a function of  \inflation, for that part of the evolution where $Y_{\rm s}<0.3$. The colour bar indicates the effective temperature of the models.}
\label{appfig:conveff}
\end{figure}

The maximum value of the opacity around the Fe-bump ($\kmax$; Fig.\,\ref{appfig:kappa}) and the density at the location of $\kmax$ (Fig.\,\ref{appfig:dens}) decrease initially as \inflation increases, because of an increase in $L_{\rm rad}/M$ (Fig.\,\ref{appfig:Lrad}). The product of the quantities $\kmax$ and $L_{\rm rad}/M$, which is proportional to $\Gamma$, also increases initially up to \inflation $\approx0.1$ (Fig.\,\ref{appfig:gamma}).  Thereafter it starts decreasing with a decrease in $L_{\rm rad}/M$.  The $L_{\rm rad}/M$ decreases in this phase of the evolution because of a rise in convective efficiency at this location, shown in Fig.\,\ref{appfig:conveff}. This is because as $\teff$ of the model keeps decreasing, the Fe-bump moves deeper inside the star where density is higher and hence convection is relatively efficient. When convection is capable of transporting the energy, the radiative luminosity $L_{\rm rad}$ decreases and hence, the Eddington factor at this location ($\Gamma_{\rm Fe}$) decreases. The inflated envelope keeps on increasing in size in spite of $\Gamma_{\rm Fe}$ coming down to values as low as $0.91$. 

\begin{figure}
\centering
\mbox{\includegraphics[width=6.5cm,angle=-90]{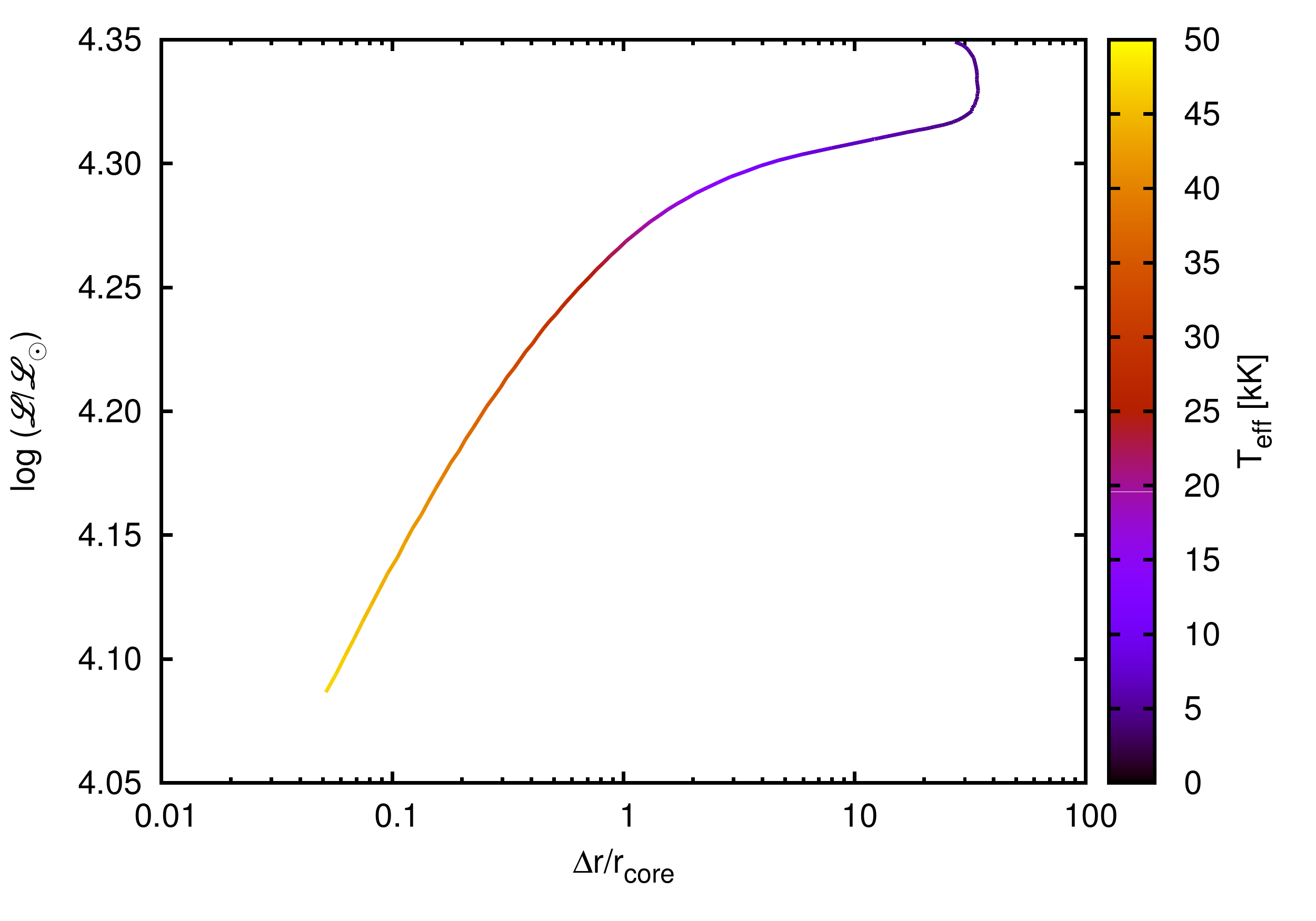}}
   \caption{Evolution of the quantity $\log(\shrL)$ as a function of \inflation for that part of the evolution where $Y_{\rm s}<0.3$. The colour bar indicates the effective temperature of the models.}
\label{appfig:LM}
\end{figure}

At \inflation$\gtrsim20$, $L_{\rm rad}/M$ and $\Gamma_{\rm Fe}$ increase again while $\kmax$ decreases. Since the star experiences high mass-loss rates at such low effective temperatures, its $L/M$ ratio increases sharply in this phase (Fig.\,\ref{appfig:LM}), but the convective efficiency does not increase as much. Hence
to let the relatively high radiative flux to pass through, the model reduces its opacity which pushes up the value of $\Gamma$ at that location. Note that at $\teff$ below $\sim 8000\,$K the location of $\gammax$ is in the hydrogen recombination zone and not within the Fe-bump.

\end{document}